\documentclass[%
reprint,
superscriptaddress,
 amsmath,amssymb,
 aip,
apl
]{revtex4-1}

\usepackage{graphicx}
\usepackage{dcolumn}
\usepackage{bm}
\usepackage{upgreek} 
\usepackage{times}

\begin{document}

\title{Enhanced magnon spin transport in NiFe$_2$O$_4$ thin films on a lattice-matched substrate}


\author{J. Shan}
\email[]{j.shan@rug.nl}
\affiliation{Physics of Nanodevices, Zernike Institute for Advanced Materials, University of Groningen, Nijenborgh 4, 9747 AG Groningen, The Netherlands}
\author{A. V. Singh}
\affiliation{Center for Materials for Information Technology, The University of Alabama, Tuscaloosa, AL 35487, USA}
\author{L. Liang}
\author{L. Cornelissen}
\affiliation{Physics of Nanodevices, Zernike Institute for Advanced Materials, University of Groningen, Nijenborgh 4, 9747 AG Groningen, The Netherlands}
\author{A. Gupta}
\affiliation{Center for Materials for Information Technology, The University of Alabama, Tuscaloosa, AL 35487, USA}
\author{B. J. van Wees}
\affiliation{Physics of Nanodevices, Zernike Institute for Advanced Materials, University of Groningen, Nijenborgh 4, 9747 AG Groningen, The Netherlands}
\author{T. Kuschel}
\affiliation{Physics of Nanodevices, Zernike Institute for Advanced Materials, University of Groningen, Nijenborgh 4, 9747 AG Groningen, The Netherlands}
\affiliation{Center for Spinelectronic Materials and Devices, Department of Physics, Bielefeld University, Universit\"{a}tsstra{\ss}e 25, 33615 Bielefeld, Germany}
\date{\today}

\begin{abstract}

We investigate magnon spin transport in epitaxial nickel ferrite (NiFe$_2$O$_4$, NFO) films grown on magnesium gallate spinel (MgGa$_2$O$_4$, MGO) substrates, which have a lattice mismatch with NFO as small as 0.78\%, resulting in the reduction of antiphase boundary defects and thus in improved magnetic properties in the NFO films. In the nonlocal transport experiments, enhanced signals are observed for both electrically and thermally excited magnons, and the magnon relaxation length ($\lambda_m$) of NFO is found to be around 2.5 $\mu$m at room temperature. Moreover, at both room and low temperatures, we present distinct features from the nonlocal spin Seebeck signals which arise from magnon-polaron formation. Our results demonstrate excellent magnon transport properties (magnon spin conductivity, $\lambda_m$ and spin mixing conductance at the interface between Pt) of NFO films grown on a lattice-matched substrate that are comparable with those of yttrium iron garnet.

\end{abstract}


\maketitle
Magnons, the collective excitation of spins, are playing the central role in the field of insulator spintronics.\cite{chumak_magnon_2015} Magnons in magnetic materials can interact with conduction electrons in adjacent heavy metals, transferring spin angular momentum and thus allowing for magnonic spin current excitation and detection using electrical methods.\cite{flipse_observation_2014,cornelissen_long-distance_2015,goennenwein_non-local_2015,li_observation_2016,wu_observation_2016,cornelissen_magnon_2016,das_spin_2017,althammer_pure_2018} Besides, magnons can be driven thermally, known as the spin Seebeck effect (SSE).\cite{uchida_observation_2008,uchida_spin_2010,bauer_spin_2012,xiao_theory_2010,kehlberger_length_2015} Both magnons generated by a spin voltage bias and a temperature gradient can be transported for a certain distance in the order of a few to tens of micrometers, as reported recently in ferrimagnetic\cite{cornelissen_long-distance_2015,shan_nonlocal_2017} and even in antiferromagnetic materials,\cite{lebrun_electrically_2018} making magnons promising as novel information carriers.

Nickel ferrite (NFO) is a ferrimagnetic insulator with inverse spinel structure. It is widely used in high-frequency systems and as inductors in conventional applications.\cite{kittel_introduction_2004} Recently, NFO and other spinel ferrites were explored for spintronics applications, where effects like spin Hall magnetoresistance (SMR),\cite{nakayama_spin_2013,althammer_quantitative_2013,isasa_spin_2014,isasa_spin_2016,ding_spin_2014} SSE\cite{meier_thermally_2013,guo_thermal_2016,niizeki_observation_2015,ramos_observation_2013,uchida_longitudinal_2010,aqeel_spin-hall_2015,bougiatioti_quantitative_2017-1,kuschel_static_2016} and nonlocal magnon spin transport\cite{shan_nonlocal_2017} were reported. In most of these studies, large magnetic fields of a few teslas are required to align the magnetization of the ferrites, possibly due to the presence of antiphase boundaries.\cite{margulies_origin_1997}

However, it was recently shown that the NFO films grown on nearly-lattice-matched substrates with similar spinel structures, such as MgGa$_2$O$_4$ and CoGa$_2$O$_4$, exhibited superior magnetic properties due to the elimination of antiphase boundaries, leading to, for instance, a larger saturation magnetization ($M_{\text{S}}$), smaller coercive fields and a lower Gilbert damping constant, compared to the NFO films grown on the typically used MgAl$_2$O$_4$ (MAO) substrate.\cite{singh_bulk_2017} An enhanced longitudinal SSE effect was reported on such NFO films.\cite{rastogi_enhancement_nodate} It can be expected that the nonlocal transport properties of magnon spin are also elevated in these NFO films, as we discuss in this paper.

\begin{figure*}
	\includegraphics[width=18cm]{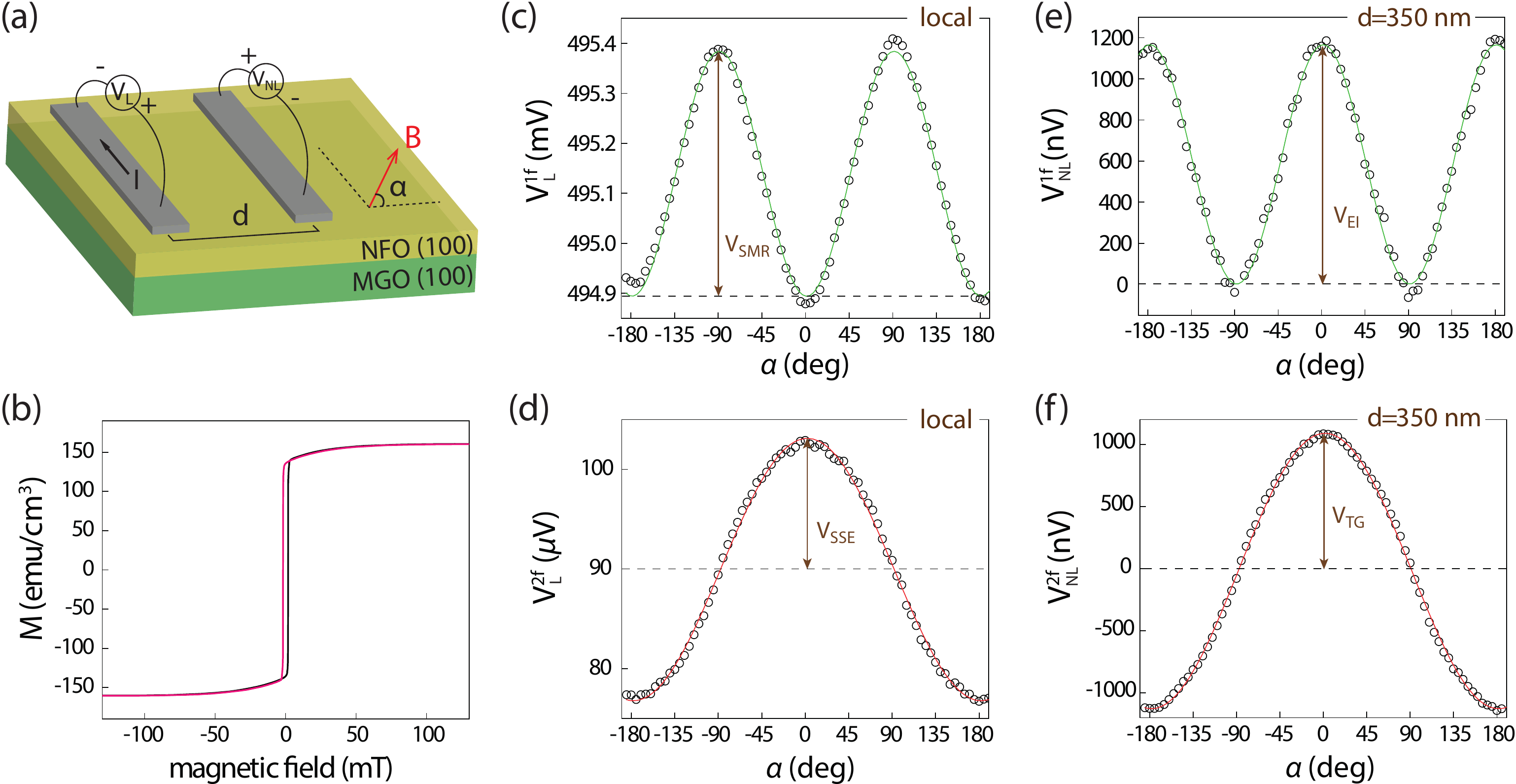}
	\caption{(a) Schematic geometry of local and nonlocal measurements. An electric current $I$ is applied at one Pt strip, and voltages can be detected at the same strip (locally) or at the other one (nonlocally). An in-plane magnetic field is applied at an angle denoted by $\alpha$. 
	(b) In-plane magnetization of the 450 nm-thick NFO film obtained from SQUID at room temperature. (c)-(f), Room-temperature local and nonlocal measurements shown in first and second harmonic signals, with $I=100 \ \mu$A. They are measured on the 40-nm-thick NFO sample with an external magnetic field of 300 mT under angular sweep. Only for (e), a background of 910 nV is subtracted.}
	\label{angular}
\end{figure*}

We studied two NFO films on MGO (100) substrates, with thicknesses of 40 nm and 450 nm, respectively. NFO films were grown by pulsed laser deposition, in the same way as described in Refs.~\cite{singh_bulk_2017,rastogi_enhancement_nodate}. Prior to further processes, the 450-nm-thick sample was characterized by superconducting quantum interference device (SQUID) magnetometry, exhibiting an in-plane coercive field lower than 5 mT (see Fig.~\ref{angular}(b)). Afterwards, multiple devices were fabricated on both samples. Figure \ref{angular}(a) shows schematically the typical geometry of a device, where two identical Pt strips are patterned in parallel with a center-to-center spacing $d$, ranging from 0.3 to 25 $\mu$m for all devices. The lengths and widths of the Pt strips are designed to be different for shorter- and longer-$d$ devices, as summarized in Table \ref{table:geometry}. In Geometry I, Pt strips are 100 nm in width, allowing for fabrication of devices with narrow spacings. In Geometry II, Pt strips are wider and longer, permitting larger injection currents which yield larger signal-to-noise ratio, so that small signals can be resolved. For all devices, Pt is sputtered with a thickness of 8 nm, showing a conductivity of around 3 $\times$10$^6$ S/m. Contacts consisting of Ti (5 nm)/Au (60 nm) were patterned in the final step of device fabrication. 

\begin{table}[b] 
	\caption{Sample details of Geometry I and II.}
	\begin{ruledtabular}
		\begin{tabular} {c c c c}
			Geometry & Pt length ($\mu$m) & Pt width ($\mu$m) & distances ($\mu$m) \\
			\hline
			I & 10 & 0.1 & 0.3 - 2\\
			II & 20 & 0.5 & 2 - 25\\
		\end{tabular}
	\end{ruledtabular}
	\label{table:geometry}
\end{table}

Electrical measurements were performed with a standard lock-in technique, where a low-frequency ac current, $I= \sqrt 2 I_0 \sin (2 \pi ft)$, was used as the input to the device, and voltage outputs were detected at the same ($1f$) or double frequency ($2f$), representing the linear and quadratic effects, respectively. In this study, typically $I_0$  is 100 $\mu$A and $f$ is set to be around 13 Hz. For the local detection $V_{\text{L}}$, as shown in Fig.~\ref{angular}(a), $V_{\text{L}}^{1f}$ detects the resistance and magnetoresistance (MR) effect of the Pt strip, and $V_{\text{L}}^{2f}$ incorporates the current-induced local SSE. \cite{schreier_current_2013,vlietstra_simultaneous_2014} For the nonlocal detection $V_{\text{NL}}$, $V_{\text{NL}}^{1f}$ represents the nonlocal signals from magnons that are injected electrically via SHE,\cite{cornelissen_long-distance_2015,goennenwein_non-local_2015} and  $V_{\text{NL}}^{2f}$ stands for the nonlocal SSE.\cite{cornelissen_long-distance_2015,shan_influence_2016,giles_long-range_2015,shan_nonlocal_2017,shan_criteria_2017,giles_thermally_2017,cornelissen_nonlocal_2017} The conductance of the NFO thin films was checked by measuring resistances between random pairs of electrically detached contacts, which yielded values over G$\Omega$, confirming the insulating nature of the NFO films.

We first perform angular-dependent measurements at room temperature for both local and nonlocal configurations, with results plotted in Figs.~\ref{angular}(c)-(f). The sample was rotated in-plane with a constant magnetic field applied. The strength of the field is 300 mT, large enough to saturate the NFO magnetization along the field direction. A strong MR effect, with $\Delta R/R \approx 0.1 \%$, was observed from the local $V_{\text{L}}^{1f}$ signal (see Fig.~\ref{angular}(c)). This MR effect was checked to be magnetic-field independent in the range from 100 to 400 mT, indicating that the observed MR effect is the SMR effect which is sensitive to the NFO magnetization that is saturated in this range, instead of the Hanle MR effect \cite{velez_hanle_2016} which depends on the external magnetic field. This is in marked contrast to the previous observations from sputtered NFO thin films grown on MAO, where only the Hanle MR effect was observed at fields above 1 T.\cite{shan_nonlocal_2017} The SMR ratios for both 40- and 450-nm thick samples exhibit similar values, ranging between 0.07\% to 0.1\%, around 3 to 4 times larger than those for Pt/yttrium iron garnet (YIG) systems with similar Pt thickness.\cite{vlietstra_spin-hall_2013,cornelissen_magnon_2016,shan_influence_2016} It is also more than twice as large as the SMR reported from Pt/NFO systems with the NFO layer grown by chemical vapor deposition on MAO substrates.\cite{althammer_quantitative_2013} Using the average SMR ratio of 0.08\% and the spin Hall angle of Pt of 0.11,\cite{cornelissen_magnon_2016,shan_influence_2016} we estimated the real part of the spin mixing conductance ($G_r$) for Pt/NFO systems to be $5.7 \times 10^{14}$ S/m$^2$ with the SMR equation,\cite{chen_theory_2013} being more than 3 times larger than that of the Pt/YIG systems determined with the same method.\cite{cornelissen_magnon_2016}

 \begin{figure}
	\includegraphics[width=8cm]{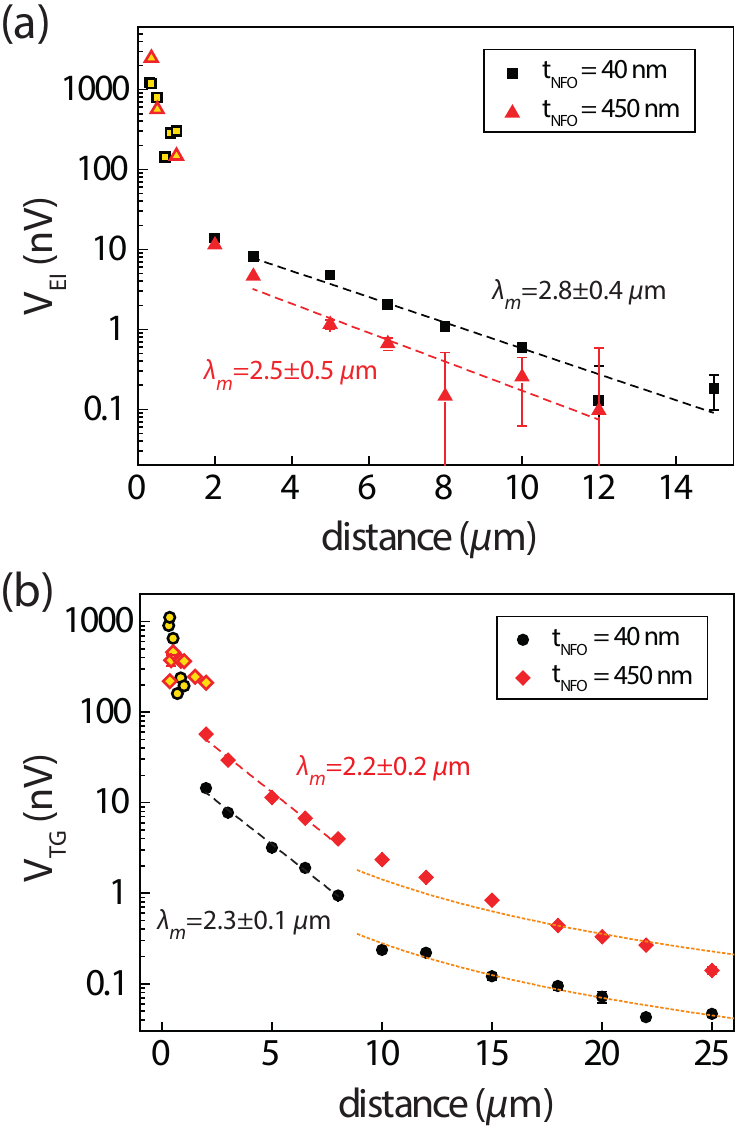}
	\caption{Distance dependence of (a) $V_{\text{EI}}$ and (b) $V_{\text{TG}}$ measured at B=200 mT on both NFO samples at room temperature, normalized to $I=$100 $\mu$A. The datapoints filled with yellow color are obtained from devices in Geometry I, while the rest belongs to Geometry II. The datapoints from Geometry II are normalized to Geometry I as described in Ref.~\cite{shan_influence_2016} for better comparison.  Dashed lines are exponential fittings with the formula $V=A \exp(-d/ \lambda_m)$, with the coefficient $A$ being different for each fitting. The extracted $\lambda_m$ from each fitting is indicated nearby. The dotted orange lines in (b) are $1/d^2$ fittings for long-$d$ results.}
	\label{exponential} 
\end{figure}

Figures \ref{angular}(e) and \ref{angular}(f) plot typical results from the nonlocal measurements in $V_{\text{NL}}^{1f}$ and $V_{\text{NL}}^{2f}$, showing respectively $\cos^2(\alpha)$ and $\cos (\alpha)$ dependences, same as observed previously in YIG or NFO films with Pt or Ta electrodes.\cite{cornelissen_long-distance_2015,shan_influence_2016,shan_nonlocal_2017,liu_magnon_2017,liu_nonlocal_2018} For the magnon transport process represented by $V_{\text{NL}}^{1f}$, both the magnon excitation and detection efficiencies are governed by $\cos (\alpha)$, which in total yields a $\cos^2(\alpha)$ behavior. For $V_{\text{NL}}^{2f}$, on the other hand, the thermal magnon excitation is independent of $\alpha$ but the detection process is, thus showing a $\cos (\alpha)$ dependence. Their amplitudes, denoted as $V_{\text{EI}}$ and $V_{\text{TG}}$ respectively, can be obtained from sinusoidal fittings. 

Next, we present $V_{\text{EI}}$ and $V_{\text{TG}}$ for all devices as a function of $d$ on both the 40- and 450-nm-thick samples to investigate the magnon relaxation properties, as shown in Fig.~\ref{exponential}. For both $V_{\text{EI}}$ and $V_{\text{TG}}$, discontinuities are found between Geometry I ($d \leq 2 \ \mu$m, filled with yellow color) and II ($d \geq 2 \ \mu$m), even though the data from Geometry II are carefully normalized to Geometry I as was done for Pt/YIG nonlocal devices to link the data between the two geometries. \cite{shan_influence_2016} However, this normalization method is based on the assumption of noninvasive contacts and does not account for the additional spin absorption that was induced by widening the Pt contact width. This normalization method works well for Pt/YIG systems but becomes less satisfactory for Pt/NFO systems as we study here, which is expected in view of a larger $G_r$ value. 

For $V_{\text{EI}}$, the datapoints at $d>$15 $\mu$m ($d>$ 12 $\mu$m for 450 nm NFO) are not plotted as the signal amplitudes become much smaller than the noise level. For shorter distances ($d<$1 $\mu$m), the signals on both samples are even comparable to those measured on thin YIG films with similar device geometry \cite{cornelissen_long-distance_2015,shan_influence_2016}, though a fairer comparison should be made with the same thickness of the magnetic insulators. We can also make a comparison between the $V_{\text{EI}}$ signals from the 40-nm-thick NFO film studied here and the 44-nm-thick sputtered NFO film on MAO substrate studied in Ref.~\cite{shan_nonlocal_2017}. We found that for the same device geometry ($d=$350 nm) and Pt thickness, the $V_{\text{EI}}$ signal amplitudes obtained here is around 100 times larger than found in  Ref.~\cite{shan_nonlocal_2017}, showing the superior quality of the NFO films studied in this paper.

To extract $\lambda_m$ for these NFO samples at room temperature, we performed exponential fittings as shown in Fig.~\ref{exponential}(a) by the dashed lines. We limit the fit to the datapoints in the exponential regime where $d>2 \ \mu$m. Both datasets yield $\lambda_m \approx$ 2.5 $\mu$m for the two NFO samples with different thicknesses. Worthnotingly,  the $V_{\text{EI}}$ signals for the 450 nm NFO are in general smaller than those for the 40 nm NFO sample, except for one datapoint at the shortest distance. However, one would expect the opposite, as increasing the NFO thickness from 40 to 450 nm enlarges the magnon conductance without introducing extra relaxation channel vertically, given that 450 nm is still much smaller than $\lambda_m \approx$ 2.5 $\mu$m. This puzzle is similar as for Pt/YIG systems \cite{shan_influence_2016} and the reason is not yet clear to us. 

Now we move to the thermally generated nonlocal SSE signals $V_{\text{TG}}$ as shown in Fig.~\ref{exponential}(b). 
According to the bulk-generated SSE picture,\cite{duine_universal_2017,shan_influence_2016,cornelissen_magnon_2016,cornelissen_nonlocal_2017} at a certain distance ($d_{\text{rev}}$) $V_{\text{TG}}$ should reverse sign, where in short distances $V_{\text{TG}}$ has the same sign as the local SSE signal, and further away the sign alters. $d_{\text{rev}}$ is influenced by the thickness of the magnetic insulator and interfacial spin transparency at the  contacts.\cite{shan_influence_2016} 
With our measurement configuration (the polarities of local and nonlocal measurement configurations are opposite as shown in Fig.~\ref{angular}(a)), the $V_{\text{TG}}$ measured from all devices are in fact opposite in the sign compared to the local SSE signals (see Fig.~\ref{angular}(d)), meaning $d_{\text{rev}}$ is positioned closer than the shortest $d$ we investigated. Only an upturn is observable for $V_{\text{TG}}$ of 450 nm NFO sample at short-$d$ range. Compared to Pt/YIG systems, where $d_{\text{rev}}$ is about 1.6 times of the YIG thickness, for Pt/NFO systems the sign-reversal takes place much closer to the heater, possibly because of the Pt/NFO interface being more transparent for a larger $G_r$. 

Exponential fittings can also be carried out for $V_{\text{TG}}$ on both samples. Note that only the datapoints in the exponential regime can be used to extract $\lambda_m$, which typical starts at $d=\lambda_m$ and extends to a few $\lambda_m$.\cite{shan_criteria_2017} Further than the exponential regime, $V_{\text{TG}}$ starts to decay geometrically as $1/d^2$, dominated by the temperature gradient present near the detector. Based on the $\lambda_m$ that we extracted from the decay of the electrically injected magnon signals, we identify $2 \leq d \leq 8 \ \mu$m as the exponential regime and obtained $\lambda_m$ to be around 2.2 or 2.3 $\mu$m from the decay of $V_{\text{TG}}$. The consistency between the $\lambda_m$ found from magnon signals excited electrically and thermally illustrate again the same transport nature of the magnons generated in both methods.

\begin{figure}[t]
	\includegraphics[width=8cm]{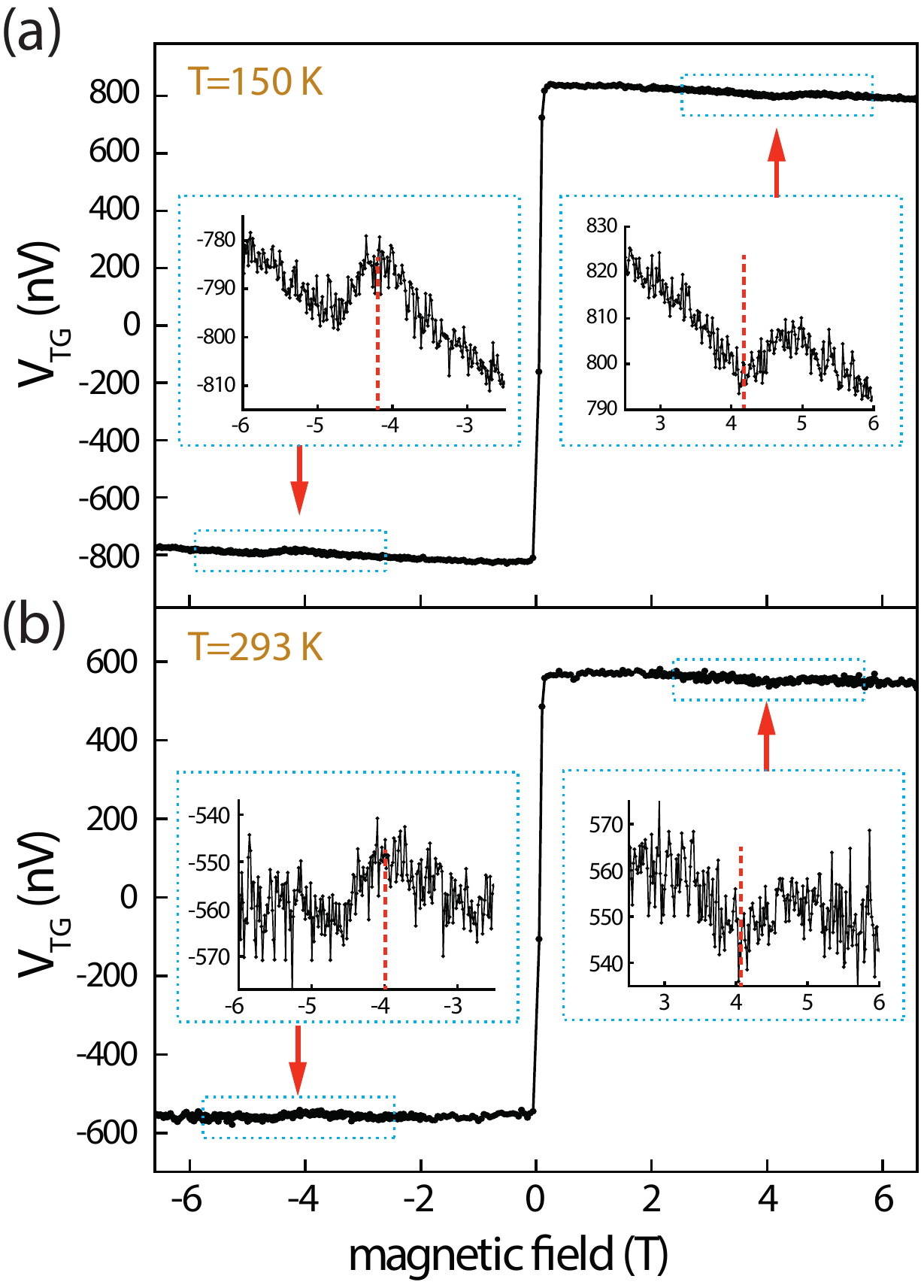}
	\caption{Magnetic field sweep measurements of $V_{\text{TG}}$ at (a) $T=150$ K and (b) $T=293$ K, on the 450 nm NFO sample ($d$=1 $\mu$m) from the nonlocal spin Seebeck measurements with $\alpha=0^{\circ}$. Insets show close-ups of the resonant dips.}
	\label{polaron} 
\end{figure}

Owing to the excellent quality of the NFO films, we are able to study its magnetoelastic coupling by means of the nonlocal SSE. It was observed in YIG that for both the local and nonlocal SSE signals, spike structures arose at certain magnetic fields, at which the magnon and phonon dispersions became tangent to each other, resulting in maximal magnetoelastic interaction and the formation of magnon-polarons. \cite{kikkawa_magnon_2016,flebus_magnon-polaron_2017,cornelissen_nonlocal_2017,man_direct_2017} At these conditions, the spin Seebeck signals have extra contributions from the magnon-polarons, provided that the magnon and phonon impurity scattering potentials are different.\cite{kikkawa_magnon_2016,flebus_magnon-polaron_2017} It was found that for YIG films, the acoustic quality is higher than the magnetic one, with \textit{peaks} observed in local SSE and nonlocal SSE ($d<d_{\text{rev}}$) measurements and \textit{dips} observed for nonlocal SSE where $d>d_{\text{rev}}$.\cite{kikkawa_magnon_2016,cornelissen_nonlocal_2017} This effect is explained as several parameters such as  $\lambda_m$, the bulk spin Seebeck coefficient, the magnon spin and heat conductivities are all modified by the emergence of magnon-polarons.\cite{flebus_magnon-polaron_2017,cornelissen_nonlocal_2017} So far, this resonant enhancement/suppression of SSE caused by magnetoelastic coupling has only been clearly observed in YIG; besides, a bimodal structure was found in the SSE of a Ni$_{0.65}$Zn$_{0.35}$Al$_{0.8}$Fe$_{1.2}$O$_{4}$ thin film and was speculated to be related to magnon-phonon interactions.\cite{wang_bimodal_2018}

Here we present distinctive magnon-polaron features in the nonlocal SSE measurements on our NFO films. Figure \ref{polaron} shows field-sweep data of $V_{\text{TG}}$ performed on one device of the 450-nm-thick NFO sample at $T=$ 150 K and 293 K. At both temperatures, asymmetric dip structures of $V_{\text{TG}}$ are clearly visible, around $\pm$4.2 T at $T=$150 K and shift to $\pm$4.0 T at $T=$293 K. The change of the characteristic magnetic field of 0.2 T for a temperature decrease of about 150 K is comparable to Pt/YIG systems.\cite{cornelissen_nonlocal_2017} The sign of the anomalies is in accordance with the previous observation reported in Ref.~\cite{cornelissen_nonlocal_2017}, considering that the spacing between the Pt strips ($d=$1 $\mu$m) is further than $d_{\text{rev}}$. This implies that the studied NFO film may also have a higher acoustic than magnetic quality like YIG, though a careful study which measures the anomalies from the local SSE is needed. 

The magnetic fields where the anomalies occur can be evaluated by the phonon and magnon dispersions. In our experiments, limited by the maximal applied magnetic field ($\mu_0H \approx 7$ T), we could only probe the first anomaly which involves transverse acoustic (TA) phonons with a lower sound velocity. The TA phonons follow the dispersion relation $\omega=v_{\text{T}}k$, where $v_{\text{T}}$ is the TA phonon sound velocity. $v_{\text{T}}$ is related to the elastic constant $C_{44}$ and material density $\rho$ by $C_{44}=\rho v_{\text{T}}^2$, \cite{wang_bimodal_2018,gulseren_high-pressure_2002} and is determined to be 3968 m/s for NFO using $C_{44}=$82.3 GPa and $\rho=5230$ kg/m$^3$.\cite{li_single_1990,rastogi_enhancement_nodate} 

We assume that magnons in NFO can be also described by a parabolic dispersion relation like for YIG ($\omega=\sqrt{(D_{\text{ex}}k^2+\gamma\mu_0H)(D_{\text{ex}}k^2+\gamma(\mu_0H+M_{\text{S}}))}$), where $D_{\text{ex}}$ is the exchange stiffness, $\mu_0$ the vacuum permeability and $\gamma$ the gyromagnetic ratio. From Fig.~\ref{angular}(b) we obtain $M_{\text{S}}$ of our NFO sample to be 160 emu/cm$^3$ at room temperature, which equals 201 mT. To the best of our knowledge, the $D_{\text{ex}}$ of NFO is not experimentally reported. In our experiment, from the peak positions observed at room temperature ($\mu_0H_{\text{TA}}=\pm 4.0$ T) we can determine the only unknown parameter $D_{\text{ex}}$ to be 5.5$\times$10$^{-6}$ m$^2$/s with both phonon and magnon dispersions. This value is close to the $D_{\text{ex}}$ which can be estimated from the exchange integrals among Ni$^{2+}$, Fe$^{3+}$ (octahedral site) and Fe$^{3+}$ (tetrahedral site). \cite{srivastava_spin_1987,franco_spin-wave_2015} Using the parameters given in Ref.~\cite{kodama_surface_1996}, $D_{\text{ex}}$ of NFO can be estimated to be 6.4$\times$10$^{-6}$ m$^2$/s, within 17\% difference of our experimental value.

Anomalies were also observed in the 450 nm NFO sample for electrically excited magnons in the field-sweep measurements of $V_{\text{EI}}$ at $T=$150 K, albeit with a lower signal-to-noise ratio. For the 40 nm NFO sample, however, no clear anomalies were identified in the measured range ($\mu_0H \leq 6.6$ T) for $V_{\text{EI}}$ or $V_{\text{TG}}$. 

In summary, we have studied the magnon spin transport properties of epitaxial NFO films grown on MGO substrates in a nonlocal geometry. We obtained large nonlocal signals for both electrically and thermally excited magnons at short contact spacings, comparable to that of YIG. From the relaxation regime $\lambda_m$ was found to be around 2.5 $\mu$m. Further, we observed anomalous features as a result of magnon-polarons formation in the field-dependent SSE measurements at both 150 and 293 K, from which the exchange stiffness constant of NFO can be determined. Our results demonstrate the improved quality of NFO grown on a lattice-matched substrate, showing NFO to be a potential alternative to YIG for spintronic applications.

We thank the helpful discussions with Gerrit Bauer, Matthias Althammer and Koichi Oyanagi, and would like to acknowledge M. de Roosz, H. Adema, T. Schouten and J. G. Holstein for technical assistance. This work is supported by the research programs ``Magnon Spintronics'' and ``Skyrmionics'' of the Netherlands Organisation for Scientific Research (NWO), the NWO Spinoza prize awarded to Prof.~B.~J.~van Wees, DFG Priority Programme 1538 ``Spin Caloric Transport'' (KU 3271/1-1), NanoLab NL, EU FP7 ICT Grant InSpin 612759 and the Zernike Institute for Advanced Materials.


%

\vspace{1cm}


\begin{thebibliography}{55}%
	\makeatletter
	\providecommand \@ifxundefined [1]{%
		\@ifx{#1\undefined}
	}%
	\providecommand \@ifnum [1]{%
		\ifnum #1\expandafter \@firstoftwo
		\else \expandafter \@secondoftwo
		\fi
	}%
	\providecommand \@ifx [1]{%
		\ifx #1\expandafter \@firstoftwo
		\else \expandafter \@secondoftwo
		\fi
	}%
	\providecommand \natexlab [1]{#1}%
	\providecommand \enquote  [1]{``#1''}%
	\providecommand \bibnamefont  [1]{#1}%
	\providecommand \bibfnamefont [1]{#1}%
	\providecommand \citenamefont [1]{#1}%
	\providecommand \href@noop [0]{\@secondoftwo}%
	\providecommand \href [0]{\begingroup \@sanitize@url \@href}%
	\providecommand \@href[1]{\@@startlink{#1}\@@href}%
	\providecommand \@@href[1]{\endgroup#1\@@endlink}%
	\providecommand \@sanitize@url [0]{\catcode `\\12\catcode `\$12\catcode
		`\&12\catcode `\#12\catcode `\^12\catcode `\_12\catcode `\%12\relax}%
	\providecommand \@@startlink[1]{}%
	\providecommand \@@endlink[0]{}%
	\providecommand \url  [0]{\begingroup\@sanitize@url \@url }%
	\providecommand \@url [1]{\endgroup\@href {#1}{\urlprefix }}%
	\providecommand \urlprefix  [0]{URL }%
	\providecommand \Eprint [0]{\href }%
	\providecommand \doibase [0]{http://dx.doi.org/}%
	\providecommand \selectlanguage [0]{\@gobble}%
	\providecommand \bibinfo  [0]{\@secondoftwo}%
	\providecommand \bibfield  [0]{\@secondoftwo}%
	\providecommand \translation [1]{[#1]}%
	\providecommand \BibitemOpen [0]{}%
	\providecommand \bibitemStop [0]{}%
	\providecommand \bibitemNoStop [0]{.\EOS\space}%
	\providecommand \EOS [0]{\spacefactor3000\relax}%
	\providecommand \BibitemShut  [1]{\csname bibitem#1\endcsname}%
	\let\auto@bib@innerbib\@empty
	\bibitem [{\citenamefont {Chumak}\ \emph {et~al.}(2015)\citenamefont {Chumak},
		\citenamefont {Vasyuchka}, \citenamefont {Serga},\ and\ \citenamefont
		{Hillebrands}}]{chumak_magnon_2015}%
	\BibitemOpen
	\bibfield  {author} {\bibinfo {author} {\bibfnamefont {A.~V.}\ \bibnamefont
			{Chumak}}, \bibinfo {author} {\bibfnamefont {V.~I.}\ \bibnamefont
			{Vasyuchka}}, \bibinfo {author} {\bibfnamefont {A.~A.}\ \bibnamefont
			{Serga}}, \ and\ \bibinfo {author} {\bibfnamefont {B.}~\bibnamefont
			{Hillebrands}},\ }\href {\doibase 10.1038/nphys3347} {\bibfield  {journal}
		{\bibinfo  {journal} {Nature Physics}\ }\textbf {\bibinfo {volume} {11}},\
		\bibinfo {pages} {453} (\bibinfo {year} {2015})}\BibitemShut {NoStop}%
	\bibitem [{\citenamefont {Flipse}\ \emph {et~al.}(2014)\citenamefont {Flipse},
		\citenamefont {Dejene}, \citenamefont {Wagenaar}, \citenamefont {Bauer},
		\citenamefont {Youssef},\ and\ \citenamefont {van
			Wees}}]{flipse_observation_2014}%
	\BibitemOpen
	\bibfield  {author} {\bibinfo {author} {\bibfnamefont {J.}~\bibnamefont
			{Flipse}}, \bibinfo {author} {\bibfnamefont {F.~K.}\ \bibnamefont {Dejene}},
		\bibinfo {author} {\bibfnamefont {D.}~\bibnamefont {Wagenaar}}, \bibinfo
		{author} {\bibfnamefont {G.~E.~W.}\ \bibnamefont {Bauer}}, \bibinfo {author}
		{\bibfnamefont {J.~B.}\ \bibnamefont {Youssef}}, \ and\ \bibinfo {author}
		{\bibfnamefont {B.~J.}\ \bibnamefont {van Wees}},\ }\href {\doibase
		10.1103/PhysRevLett.113.027601} {\bibfield  {journal} {\bibinfo  {journal}
			{Physical Review Letters}\ }\textbf {\bibinfo {volume} {113}},\ \bibinfo
		{pages} {027601} (\bibinfo {year} {2014})}\BibitemShut {NoStop}%
	\bibitem [{\citenamefont {Cornelissen}\ \emph {et~al.}(2015)\citenamefont
		{Cornelissen}, \citenamefont {Liu}, \citenamefont {Duine}, \citenamefont
		{Youssef},\ and\ \citenamefont {van Wees}}]{cornelissen_long-distance_2015}%
	\BibitemOpen
	\bibfield  {author} {\bibinfo {author} {\bibfnamefont {L.~J.}\ \bibnamefont
			{Cornelissen}}, \bibinfo {author} {\bibfnamefont {J.}~\bibnamefont {Liu}},
		\bibinfo {author} {\bibfnamefont {R.~A.}\ \bibnamefont {Duine}}, \bibinfo
		{author} {\bibfnamefont {J.~B.}\ \bibnamefont {Youssef}}, \ and\ \bibinfo
		{author} {\bibfnamefont {B.~J.}\ \bibnamefont {van Wees}},\ }\href {\doibase
		10.1038/nphys3465} {\bibfield  {journal} {\bibinfo  {journal} {Nature
				Physics}\ }\textbf {\bibinfo {volume} {11}},\ \bibinfo {pages} {1022}
		(\bibinfo {year} {2015})}\BibitemShut {NoStop}%
	\bibitem [{\citenamefont {Goennenwein}\ \emph {et~al.}(2015)\citenamefont
		{Goennenwein}, \citenamefont {Schlitz}, \citenamefont {Pernpeintner},
		\citenamefont {Ganzhorn}, \citenamefont {Althammer}, \citenamefont {Gross},\
		and\ \citenamefont {Huebl}}]{goennenwein_non-local_2015}%
	\BibitemOpen
	\bibfield  {author} {\bibinfo {author} {\bibfnamefont {S.~T.~B.}\
			\bibnamefont {Goennenwein}}, \bibinfo {author} {\bibfnamefont
			{R.}~\bibnamefont {Schlitz}}, \bibinfo {author} {\bibfnamefont
			{M.}~\bibnamefont {Pernpeintner}}, \bibinfo {author} {\bibfnamefont
			{K.}~\bibnamefont {Ganzhorn}}, \bibinfo {author} {\bibfnamefont
			{M.}~\bibnamefont {Althammer}}, \bibinfo {author} {\bibfnamefont
			{R.}~\bibnamefont {Gross}}, \ and\ \bibinfo {author} {\bibfnamefont
			{H.}~\bibnamefont {Huebl}},\ }\href {\doibase 10.1063/1.4935074} {\bibfield
		{journal} {\bibinfo  {journal} {Applied Physics Letters}\ }\textbf {\bibinfo
			{volume} {107}},\ \bibinfo {pages} {172405} (\bibinfo {year}
		{2015})}\BibitemShut {NoStop}%
	\bibitem [{\citenamefont {Li}\ \emph {et~al.}(2016)\citenamefont {Li},
		\citenamefont {Xu}, \citenamefont {Aldosary}, \citenamefont {Tang},
		\citenamefont {Lin}, \citenamefont {Zhang}, \citenamefont {Lake},\ and\
		\citenamefont {Shi}}]{li_observation_2016}%
	\BibitemOpen
	\bibfield  {author} {\bibinfo {author} {\bibfnamefont {J.}~\bibnamefont
			{Li}}, \bibinfo {author} {\bibfnamefont {Y.}~\bibnamefont {Xu}}, \bibinfo
		{author} {\bibfnamefont {M.}~\bibnamefont {Aldosary}}, \bibinfo {author}
		{\bibfnamefont {C.}~\bibnamefont {Tang}}, \bibinfo {author} {\bibfnamefont
			{Z.}~\bibnamefont {Lin}}, \bibinfo {author} {\bibfnamefont {S.}~\bibnamefont
			{Zhang}}, \bibinfo {author} {\bibfnamefont {R.}~\bibnamefont {Lake}}, \ and\
		\bibinfo {author} {\bibfnamefont {J.}~\bibnamefont {Shi}},\ }\href {\doibase
		10.1038/ncomms10858} {\bibfield  {journal} {\bibinfo  {journal} {Nature
				Communications}\ }\textbf {\bibinfo {volume} {7}},\ \bibinfo {pages} {10858}
		(\bibinfo {year} {2016})}\BibitemShut {NoStop}%
	\bibitem [{\citenamefont {Wu}\ \emph {et~al.}(2016)\citenamefont {Wu},
		\citenamefont {Wan}, \citenamefont {Zhang}, \citenamefont {Yuan},
		\citenamefont {Zhang}, \citenamefont {Qin}, \citenamefont {Wei},
		\citenamefont {Han},\ and\ \citenamefont {Zhang}}]{wu_observation_2016}%
	\BibitemOpen
	\bibfield  {author} {\bibinfo {author} {\bibfnamefont {H.}~\bibnamefont
			{Wu}}, \bibinfo {author} {\bibfnamefont {C.~H.}\ \bibnamefont {Wan}},
		\bibinfo {author} {\bibfnamefont {X.}~\bibnamefont {Zhang}}, \bibinfo
		{author} {\bibfnamefont {Z.~H.}\ \bibnamefont {Yuan}}, \bibinfo {author}
		{\bibfnamefont {Q.~T.}\ \bibnamefont {Zhang}}, \bibinfo {author}
		{\bibfnamefont {J.~Y.}\ \bibnamefont {Qin}}, \bibinfo {author} {\bibfnamefont
			{H.~X.}\ \bibnamefont {Wei}}, \bibinfo {author} {\bibfnamefont {X.~F.}\
			\bibnamefont {Han}}, \ and\ \bibinfo {author} {\bibfnamefont
			{S.}~\bibnamefont {Zhang}},\ }\href {\doibase 10.1103/PhysRevB.93.060403}
	{\bibfield  {journal} {\bibinfo  {journal} {Physical Review B}\ }\textbf
		{\bibinfo {volume} {93}},\ \bibinfo {pages} {060403} (\bibinfo {year}
		{2016})}\BibitemShut {NoStop}%
	\bibitem [{\citenamefont {Cornelissen}\ \emph {et~al.}(2016)\citenamefont
		{Cornelissen}, \citenamefont {Peters}, \citenamefont {Bauer}, \citenamefont
		{Duine},\ and\ \citenamefont {van Wees}}]{cornelissen_magnon_2016}%
	\BibitemOpen
	\bibfield  {author} {\bibinfo {author} {\bibfnamefont {L.~J.}\ \bibnamefont
			{Cornelissen}}, \bibinfo {author} {\bibfnamefont {K.~J.~H.}\ \bibnamefont
			{Peters}}, \bibinfo {author} {\bibfnamefont {G.~E.~W.}\ \bibnamefont
			{Bauer}}, \bibinfo {author} {\bibfnamefont {R.~A.}\ \bibnamefont {Duine}}, \
		and\ \bibinfo {author} {\bibfnamefont {B.~J.}\ \bibnamefont {van Wees}},\
	}\href {\doibase 10.1103/PhysRevB.94.014412} {\bibfield  {journal} {\bibinfo
			{journal} {Physical Review B}\ }\textbf {\bibinfo {volume} {94}},\ \bibinfo
		{pages} {014412} (\bibinfo {year} {2016})}\BibitemShut {NoStop}%
	\bibitem [{\citenamefont {Das}\ \emph {et~al.}(2017)\citenamefont {Das},
		\citenamefont {Schoemaker}, \citenamefont {van Wees},\ and\ \citenamefont
		{Vera-Marun}}]{das_spin_2017}%
	\BibitemOpen
	\bibfield  {author} {\bibinfo {author} {\bibfnamefont {K.~S.}\ \bibnamefont
			{Das}}, \bibinfo {author} {\bibfnamefont {W.~Y.}\ \bibnamefont {Schoemaker}},
		\bibinfo {author} {\bibfnamefont {B.~J.}\ \bibnamefont {van Wees}}, \ and\
		\bibinfo {author} {\bibfnamefont {I.~J.}\ \bibnamefont {Vera-Marun}},\ }\href
	{\doibase 10.1103/PhysRevB.96.220408} {\bibfield  {journal} {\bibinfo
			{journal} {Physical Review B}\ }\textbf {\bibinfo {volume} {96}},\ \bibinfo
		{pages} {220408} (\bibinfo {year} {2017})}\BibitemShut {NoStop}%
	\bibitem [{\citenamefont {Althammer}(2018)}]{althammer_pure_2018}%
	\BibitemOpen
	\bibfield  {author} {\bibinfo {author} {\bibfnamefont {M.}~\bibnamefont
			{Althammer}},\ }\href {\doibase 10.1088/1361-6463/aaca89} {\bibfield
		{journal} {\bibinfo  {journal} {Journal of Physics D: Applied Physics}\
		}\textbf {\bibinfo {volume} {51}},\ \bibinfo {pages} {313001} (\bibinfo
		{year} {2018})}\BibitemShut {NoStop}%
	\bibitem [{\citenamefont {Uchida}\ \emph {et~al.}(2008)\citenamefont {Uchida},
		\citenamefont {Takahashi}, \citenamefont {Harii}, \citenamefont {Ieda},
		\citenamefont {Koshibae}, \citenamefont {Ando}, \citenamefont {Maekawa},\
		and\ \citenamefont {Saitoh}}]{uchida_observation_2008}%
	\BibitemOpen
	\bibfield  {author} {\bibinfo {author} {\bibfnamefont {K.}~\bibnamefont
			{Uchida}}, \bibinfo {author} {\bibfnamefont {S.}~\bibnamefont {Takahashi}},
		\bibinfo {author} {\bibfnamefont {K.}~\bibnamefont {Harii}}, \bibinfo
		{author} {\bibfnamefont {J.}~\bibnamefont {Ieda}}, \bibinfo {author}
		{\bibfnamefont {W.}~\bibnamefont {Koshibae}}, \bibinfo {author}
		{\bibfnamefont {K.}~\bibnamefont {Ando}}, \bibinfo {author} {\bibfnamefont
			{S.}~\bibnamefont {Maekawa}}, \ and\ \bibinfo {author} {\bibfnamefont
			{E.}~\bibnamefont {Saitoh}},\ }\href {\doibase 10.1038/nature07321}
	{\bibfield  {journal} {\bibinfo  {journal} {Nature}\ }\textbf {\bibinfo
			{volume} {455}},\ \bibinfo {pages} {778} (\bibinfo {year}
		{2008})}\BibitemShut {NoStop}%
	\bibitem [{\citenamefont {Uchida}\ \emph
		{et~al.}(2010{\natexlab{a}})\citenamefont {Uchida}, \citenamefont {Xiao},
		\citenamefont {Adachi}, \citenamefont {Ohe}, \citenamefont {Takahashi},
		\citenamefont {Ieda}, \citenamefont {Ota}, \citenamefont {Kajiwara},
		\citenamefont {Umezawa}, \citenamefont {Kawai}, \citenamefont {Bauer},
		\citenamefont {Maekawa},\ and\ \citenamefont {Saitoh}}]{uchida_spin_2010}%
	\BibitemOpen
	\bibfield  {author} {\bibinfo {author} {\bibfnamefont {K.}~\bibnamefont
			{Uchida}}, \bibinfo {author} {\bibfnamefont {J.}~\bibnamefont {Xiao}},
		\bibinfo {author} {\bibfnamefont {H.}~\bibnamefont {Adachi}}, \bibinfo
		{author} {\bibfnamefont {J.}~\bibnamefont {Ohe}}, \bibinfo {author}
		{\bibfnamefont {S.}~\bibnamefont {Takahashi}}, \bibinfo {author}
		{\bibfnamefont {J.}~\bibnamefont {Ieda}}, \bibinfo {author} {\bibfnamefont
			{T.}~\bibnamefont {Ota}}, \bibinfo {author} {\bibfnamefont {Y.}~\bibnamefont
			{Kajiwara}}, \bibinfo {author} {\bibfnamefont {H.}~\bibnamefont {Umezawa}},
		\bibinfo {author} {\bibfnamefont {H.}~\bibnamefont {Kawai}}, \bibinfo
		{author} {\bibfnamefont {G.~E.~W.}\ \bibnamefont {Bauer}}, \bibinfo {author}
		{\bibfnamefont {S.}~\bibnamefont {Maekawa}}, \ and\ \bibinfo {author}
		{\bibfnamefont {E.}~\bibnamefont {Saitoh}},\ }\href {\doibase
		10.1038/nmat2856} {\bibfield  {journal} {\bibinfo  {journal} {Nature
				Materials}\ }\textbf {\bibinfo {volume} {9}},\ \bibinfo {pages} {894}
		(\bibinfo {year} {2010}{\natexlab{a}})}\BibitemShut {NoStop}%
	\bibitem [{\citenamefont {Bauer}, \citenamefont {Saitoh},\ and\ \citenamefont
		{van Wees}(2012)}]{bauer_spin_2012}%
	\BibitemOpen
	\bibfield  {author} {\bibinfo {author} {\bibfnamefont {G.~E.~W.}\
			\bibnamefont {Bauer}}, \bibinfo {author} {\bibfnamefont {E.}~\bibnamefont
			{Saitoh}}, \ and\ \bibinfo {author} {\bibfnamefont {B.~J.}\ \bibnamefont {van
				Wees}},\ }\href {\doibase 10.1038/nmat3301} {\bibfield  {journal} {\bibinfo
			{journal} {Nature Materials}\ }\textbf {\bibinfo {volume} {11}},\ \bibinfo
		{pages} {391} (\bibinfo {year} {2012})}\BibitemShut {NoStop}%
	\bibitem [{\citenamefont {Xiao}\ \emph {et~al.}(2010)\citenamefont {Xiao},
		\citenamefont {Bauer}, \citenamefont {Uchida}, \citenamefont {Saitoh},\ and\
		\citenamefont {Maekawa}}]{xiao_theory_2010}%
	\BibitemOpen
	\bibfield  {author} {\bibinfo {author} {\bibfnamefont {J.}~\bibnamefont
			{Xiao}}, \bibinfo {author} {\bibfnamefont {G.~E.~W.}\ \bibnamefont {Bauer}},
		\bibinfo {author} {\bibfnamefont {K.-c.}\ \bibnamefont {Uchida}}, \bibinfo
		{author} {\bibfnamefont {E.}~\bibnamefont {Saitoh}}, \ and\ \bibinfo {author}
		{\bibfnamefont {S.}~\bibnamefont {Maekawa}},\ }\href {\doibase
		10.1103/PhysRevB.81.214418} {\bibfield  {journal} {\bibinfo  {journal}
			{Physical Review B}\ }\textbf {\bibinfo {volume} {81}},\ \bibinfo {pages}
		{214418} (\bibinfo {year} {2010})}\BibitemShut {NoStop}%
	\bibitem [{\citenamefont {Kehlberger}\ \emph {et~al.}(2015)\citenamefont
		{Kehlberger}, \citenamefont {Ritzmann}, \citenamefont {Hinzke}, \citenamefont
		{Guo}, \citenamefont {Cramer}, \citenamefont {Jakob}, \citenamefont
		{Onbasli}, \citenamefont {Kim}, \citenamefont {Ross}, \citenamefont
		{Jungfleisch}, \citenamefont {Hillebrands}, \citenamefont {Nowak},\ and\
		\citenamefont {Kl{\"a}ui}}]{kehlberger_length_2015}%
	\BibitemOpen
	\bibfield  {author} {\bibinfo {author} {\bibfnamefont {A.}~\bibnamefont
			{Kehlberger}}, \bibinfo {author} {\bibfnamefont {U.}~\bibnamefont
			{Ritzmann}}, \bibinfo {author} {\bibfnamefont {D.}~\bibnamefont {Hinzke}},
		\bibinfo {author} {\bibfnamefont {E.-J.}\ \bibnamefont {Guo}}, \bibinfo
		{author} {\bibfnamefont {J.}~\bibnamefont {Cramer}}, \bibinfo {author}
		{\bibfnamefont {G.}~\bibnamefont {Jakob}}, \bibinfo {author} {\bibfnamefont
			{M.~C.}\ \bibnamefont {Onbasli}}, \bibinfo {author} {\bibfnamefont {D.~H.}\
			\bibnamefont {Kim}}, \bibinfo {author} {\bibfnamefont {C.~A.}\ \bibnamefont
			{Ross}}, \bibinfo {author} {\bibfnamefont {M.~B.}\ \bibnamefont
			{Jungfleisch}}, \bibinfo {author} {\bibfnamefont {B.}~\bibnamefont
			{Hillebrands}}, \bibinfo {author} {\bibfnamefont {U.}~\bibnamefont {Nowak}},
		\ and\ \bibinfo {author} {\bibfnamefont {M.}~\bibnamefont {Kl{\"a}ui}},\
	}\href {\doibase 10.1103/PhysRevLett.115.096602} {\bibfield  {journal}
		{\bibinfo  {journal} {Physical Review Letters}\ }\textbf {\bibinfo {volume}
			{115}},\ \bibinfo {pages} {096602} (\bibinfo {year} {2015})}\BibitemShut
	{NoStop}%
	\bibitem [{\citenamefont {Shan}\ \emph
		{et~al.}(2017{\natexlab{a}})\citenamefont {Shan}, \citenamefont
		{Bougiatioti}, \citenamefont {Liang}, \citenamefont {Reiss}, \citenamefont
		{Kuschel},\ and\ \citenamefont {van Wees}}]{shan_nonlocal_2017}%
	\BibitemOpen
	\bibfield  {author} {\bibinfo {author} {\bibfnamefont {J.}~\bibnamefont
			{Shan}}, \bibinfo {author} {\bibfnamefont {P.}~\bibnamefont {Bougiatioti}},
		\bibinfo {author} {\bibfnamefont {L.}~\bibnamefont {Liang}}, \bibinfo
		{author} {\bibfnamefont {G.}~\bibnamefont {Reiss}}, \bibinfo {author}
		{\bibfnamefont {T.}~\bibnamefont {Kuschel}}, \ and\ \bibinfo {author}
		{\bibfnamefont {B.~J.}\ \bibnamefont {van Wees}},\ }\href {\doibase
		10.1063/1.4979408} {\bibfield  {journal} {\bibinfo  {journal} {Applied
				Physics Letters}\ }\textbf {\bibinfo {volume} {110}},\ \bibinfo {pages}
		{132406} (\bibinfo {year} {2017}{\natexlab{a}})}\BibitemShut {NoStop}%
	\bibitem [{\citenamefont {Lebrun}\ \emph {et~al.}(2018)\citenamefont {Lebrun},
		\citenamefont {Ross}, \citenamefont {Bender}, \citenamefont {Qaiumzadeh},
		\citenamefont {Baldrati}, \citenamefont {Cramer}, \citenamefont {Brataas},
		\citenamefont {Duine},\ and\ \citenamefont
		{Kl{\"a}ui}}]{lebrun_electrically_2018}%
	\BibitemOpen
	\bibfield  {author} {\bibinfo {author} {\bibfnamefont {R.}~\bibnamefont
			{Lebrun}}, \bibinfo {author} {\bibfnamefont {A.}~\bibnamefont {Ross}},
		\bibinfo {author} {\bibfnamefont {S.~A.}\ \bibnamefont {Bender}}, \bibinfo
		{author} {\bibfnamefont {A.}~\bibnamefont {Qaiumzadeh}}, \bibinfo {author}
		{\bibfnamefont {L.}~\bibnamefont {Baldrati}}, \bibinfo {author}
		{\bibfnamefont {J.}~\bibnamefont {Cramer}}, \bibinfo {author} {\bibfnamefont
			{A.}~\bibnamefont {Brataas}}, \bibinfo {author} {\bibfnamefont {R.~A.}\
			\bibnamefont {Duine}}, \ and\ \bibinfo {author} {\bibfnamefont
			{M.}~\bibnamefont {Kl{\"a}ui}},\ }\href {http://arxiv.org/abs/1805.02451}
	{\bibfield  {journal} {\bibinfo  {journal} {arXiv:1805.02451 [cond-mat]}\ }
		(\bibinfo {year} {2018})},\ \bibinfo {note} {arXiv: 1805.02451}\BibitemShut
	{NoStop}%
	\bibitem [{\citenamefont {Kittel}(2004)}]{kittel_introduction_2004}%
	\BibitemOpen
	\bibfield  {author} {\bibinfo {author} {\bibfnamefont {C.}~\bibnamefont
			{Kittel}},\ }\href@noop {} {\emph {\bibinfo {title} {Introduction to {Solid}
				{State} {Physics}}}},\ \bibinfo {edition} {8th}\ ed.\ (\bibinfo  {publisher}
	{Wiley},\ \bibinfo {address} {Hoboken, NJ},\ \bibinfo {year}
	{2004})\BibitemShut {NoStop}%
	\bibitem [{\citenamefont {Nakayama}\ \emph {et~al.}(2013)\citenamefont
		{Nakayama}, \citenamefont {Althammer}, \citenamefont {Chen}, \citenamefont
		{Uchida}, \citenamefont {Kajiwara}, \citenamefont {Kikuchi}, \citenamefont
		{Ohtani}, \citenamefont {Gepr{\"a}gs}, \citenamefont {Opel}, \citenamefont
		{Takahashi}, \citenamefont {Gross}, \citenamefont {Bauer}, \citenamefont
		{Goennenwein},\ and\ \citenamefont {Saitoh}}]{nakayama_spin_2013}%
	\BibitemOpen
	\bibfield  {author} {\bibinfo {author} {\bibfnamefont {H.}~\bibnamefont
			{Nakayama}}, \bibinfo {author} {\bibfnamefont {M.}~\bibnamefont {Althammer}},
		\bibinfo {author} {\bibfnamefont {Y.-T.}\ \bibnamefont {Chen}}, \bibinfo
		{author} {\bibfnamefont {K.}~\bibnamefont {Uchida}}, \bibinfo {author}
		{\bibfnamefont {Y.}~\bibnamefont {Kajiwara}}, \bibinfo {author}
		{\bibfnamefont {D.}~\bibnamefont {Kikuchi}}, \bibinfo {author} {\bibfnamefont
			{T.}~\bibnamefont {Ohtani}}, \bibinfo {author} {\bibfnamefont
			{S.}~\bibnamefont {Gepr{\"a}gs}}, \bibinfo {author} {\bibfnamefont
			{M.}~\bibnamefont {Opel}}, \bibinfo {author} {\bibfnamefont {S.}~\bibnamefont
			{Takahashi}}, \bibinfo {author} {\bibfnamefont {R.}~\bibnamefont {Gross}},
		\bibinfo {author} {\bibfnamefont {G.~E.~W.}\ \bibnamefont {Bauer}}, \bibinfo
		{author} {\bibfnamefont {S.~T.~B.}\ \bibnamefont {Goennenwein}}, \ and\
		\bibinfo {author} {\bibfnamefont {E.}~\bibnamefont {Saitoh}},\ }\href
	{\doibase 10.1103/PhysRevLett.110.206601} {\bibfield  {journal} {\bibinfo
			{journal} {Physical Review Letters}\ }\textbf {\bibinfo {volume} {110}},\
		\bibinfo {pages} {206601} (\bibinfo {year} {2013})}\BibitemShut {NoStop}%
	\bibitem [{\citenamefont {Althammer}\ \emph {et~al.}(2013)\citenamefont
		{Althammer}, \citenamefont {Meyer}, \citenamefont {Nakayama}, \citenamefont
		{Schreier}, \citenamefont {Altmannshofer}, \citenamefont {Weiler},
		\citenamefont {Huebl}, \citenamefont {Gepr{\"a}gs}, \citenamefont {Opel},
		\citenamefont {Gross}, \citenamefont {Meier}, \citenamefont {Klewe},
		\citenamefont {Kuschel}, \citenamefont {Schmalhorst}, \citenamefont {Reiss},
		\citenamefont {Shen}, \citenamefont {Gupta}, \citenamefont {Chen},
		\citenamefont {Bauer}, \citenamefont {Saitoh},\ and\ \citenamefont
		{Goennenwein}}]{althammer_quantitative_2013}%
	\BibitemOpen
	\bibfield  {author} {\bibinfo {author} {\bibfnamefont {M.}~\bibnamefont
			{Althammer}}, \bibinfo {author} {\bibfnamefont {S.}~\bibnamefont {Meyer}},
		\bibinfo {author} {\bibfnamefont {H.}~\bibnamefont {Nakayama}}, \bibinfo
		{author} {\bibfnamefont {M.}~\bibnamefont {Schreier}}, \bibinfo {author}
		{\bibfnamefont {S.}~\bibnamefont {Altmannshofer}}, \bibinfo {author}
		{\bibfnamefont {M.}~\bibnamefont {Weiler}}, \bibinfo {author} {\bibfnamefont
			{H.}~\bibnamefont {Huebl}}, \bibinfo {author} {\bibfnamefont
			{S.}~\bibnamefont {Gepr{\"a}gs}}, \bibinfo {author} {\bibfnamefont
			{M.}~\bibnamefont {Opel}}, \bibinfo {author} {\bibfnamefont {R.}~\bibnamefont
			{Gross}}, \bibinfo {author} {\bibfnamefont {D.}~\bibnamefont {Meier}},
		\bibinfo {author} {\bibfnamefont {C.}~\bibnamefont {Klewe}}, \bibinfo
		{author} {\bibfnamefont {T.}~\bibnamefont {Kuschel}}, \bibinfo {author}
		{\bibfnamefont {J.-M.}\ \bibnamefont {Schmalhorst}}, \bibinfo {author}
		{\bibfnamefont {G.}~\bibnamefont {Reiss}}, \bibinfo {author} {\bibfnamefont
			{L.}~\bibnamefont {Shen}}, \bibinfo {author} {\bibfnamefont {A.}~\bibnamefont
			{Gupta}}, \bibinfo {author} {\bibfnamefont {Y.-T.}\ \bibnamefont {Chen}},
		\bibinfo {author} {\bibfnamefont {G.~E.~W.}\ \bibnamefont {Bauer}}, \bibinfo
		{author} {\bibfnamefont {E.}~\bibnamefont {Saitoh}}, \ and\ \bibinfo {author}
		{\bibfnamefont {S.~T.~B.}\ \bibnamefont {Goennenwein}},\ }\href {\doibase
		10.1103/PhysRevB.87.224401} {\bibfield  {journal} {\bibinfo  {journal}
			{Physical Review B}\ }\textbf {\bibinfo {volume} {87}},\ \bibinfo {pages}
		{224401} (\bibinfo {year} {2013})}\BibitemShut {NoStop}%
	\bibitem [{\citenamefont {Isasa}\ \emph {et~al.}(2014)\citenamefont {Isasa},
		\citenamefont {Bedoya-Pinto}, \citenamefont {V{\'e}lez}, \citenamefont
		{Golmar}, \citenamefont {S{\'a}nchez}, \citenamefont {Hueso}, \citenamefont
		{Fontcuberta},\ and\ \citenamefont {Casanova}}]{isasa_spin_2014}%
	\BibitemOpen
	\bibfield  {author} {\bibinfo {author} {\bibfnamefont {M.}~\bibnamefont
			{Isasa}}, \bibinfo {author} {\bibfnamefont {A.}~\bibnamefont {Bedoya-Pinto}},
		\bibinfo {author} {\bibfnamefont {S.}~\bibnamefont {V{\'e}lez}}, \bibinfo
		{author} {\bibfnamefont {F.}~\bibnamefont {Golmar}}, \bibinfo {author}
		{\bibfnamefont {F.}~\bibnamefont {S{\'a}nchez}}, \bibinfo {author}
		{\bibfnamefont {L.~E.}\ \bibnamefont {Hueso}}, \bibinfo {author}
		{\bibfnamefont {J.}~\bibnamefont {Fontcuberta}}, \ and\ \bibinfo {author}
		{\bibfnamefont {F.}~\bibnamefont {Casanova}},\ }\href {\doibase
		10.1063/1.4897544} {\bibfield  {journal} {\bibinfo  {journal} {Applied
				Physics Letters}\ }\textbf {\bibinfo {volume} {105}},\ \bibinfo {pages}
		{142402} (\bibinfo {year} {2014})}\BibitemShut {NoStop}%
	\bibitem [{\citenamefont {Isasa}\ \emph {et~al.}(2016)\citenamefont {Isasa},
		\citenamefont {V{\'e}lez}, \citenamefont {Sagasta}, \citenamefont
		{Bedoya-Pinto}, \citenamefont {Dix}, \citenamefont {S{\'a}nchez},
		\citenamefont {Hueso}, \citenamefont {Fontcuberta},\ and\ \citenamefont
		{Casanova}}]{isasa_spin_2016}%
	\BibitemOpen
	\bibfield  {author} {\bibinfo {author} {\bibfnamefont {M.}~\bibnamefont
			{Isasa}}, \bibinfo {author} {\bibfnamefont {S.}~\bibnamefont {V{\'e}lez}},
		\bibinfo {author} {\bibfnamefont {E.}~\bibnamefont {Sagasta}}, \bibinfo
		{author} {\bibfnamefont {A.}~\bibnamefont {Bedoya-Pinto}}, \bibinfo {author}
		{\bibfnamefont {N.}~\bibnamefont {Dix}}, \bibinfo {author} {\bibfnamefont
			{F.}~\bibnamefont {S{\'a}nchez}}, \bibinfo {author} {\bibfnamefont {L.~E.}\
			\bibnamefont {Hueso}}, \bibinfo {author} {\bibfnamefont {J.}~\bibnamefont
			{Fontcuberta}}, \ and\ \bibinfo {author} {\bibfnamefont {F.}~\bibnamefont
			{Casanova}},\ }\href {\doibase 10.1103/PhysRevApplied.6.034007} {\bibfield
		{journal} {\bibinfo  {journal} {Physical Review Applied}\ }\textbf {\bibinfo
			{volume} {6}},\ \bibinfo {pages} {034007} (\bibinfo {year}
		{2016})}\BibitemShut {NoStop}%
	\bibitem [{\citenamefont {Ding}\ \emph {et~al.}(2014)\citenamefont {Ding},
		\citenamefont {Chen}, \citenamefont {Liang}, \citenamefont {Zhu},
		\citenamefont {Li},\ and\ \citenamefont {Wu}}]{ding_spin_2014}%
	\BibitemOpen
	\bibfield  {author} {\bibinfo {author} {\bibfnamefont {Z.}~\bibnamefont
			{Ding}}, \bibinfo {author} {\bibfnamefont {B.~L.}\ \bibnamefont {Chen}},
		\bibinfo {author} {\bibfnamefont {J.~H.}\ \bibnamefont {Liang}}, \bibinfo
		{author} {\bibfnamefont {J.}~\bibnamefont {Zhu}}, \bibinfo {author}
		{\bibfnamefont {J.~X.}\ \bibnamefont {Li}}, \ and\ \bibinfo {author}
		{\bibfnamefont {Y.~Z.}\ \bibnamefont {Wu}},\ }\href {\doibase
		10.1103/PhysRevB.90.134424} {\bibfield  {journal} {\bibinfo  {journal}
			{Physical Review B}\ }\textbf {\bibinfo {volume} {90}},\ \bibinfo {pages}
		{134424} (\bibinfo {year} {2014})}\BibitemShut {NoStop}%
	\bibitem [{\citenamefont {Meier}\ \emph {et~al.}(2013)\citenamefont {Meier},
		\citenamefont {Kuschel}, \citenamefont {Shen}, \citenamefont {Gupta},
		\citenamefont {Kikkawa}, \citenamefont {Uchida}, \citenamefont {Saitoh},
		\citenamefont {Schmalhorst},\ and\ \citenamefont
		{Reiss}}]{meier_thermally_2013}%
	\BibitemOpen
	\bibfield  {author} {\bibinfo {author} {\bibfnamefont {D.}~\bibnamefont
			{Meier}}, \bibinfo {author} {\bibfnamefont {T.}~\bibnamefont {Kuschel}},
		\bibinfo {author} {\bibfnamefont {L.}~\bibnamefont {Shen}}, \bibinfo {author}
		{\bibfnamefont {A.}~\bibnamefont {Gupta}}, \bibinfo {author} {\bibfnamefont
			{T.}~\bibnamefont {Kikkawa}}, \bibinfo {author} {\bibfnamefont
			{K.}~\bibnamefont {Uchida}}, \bibinfo {author} {\bibfnamefont
			{E.}~\bibnamefont {Saitoh}}, \bibinfo {author} {\bibfnamefont {J.-M.}\
			\bibnamefont {Schmalhorst}}, \ and\ \bibinfo {author} {\bibfnamefont
			{G.}~\bibnamefont {Reiss}},\ }\href {\doibase 10.1103/PhysRevB.87.054421}
	{\bibfield  {journal} {\bibinfo  {journal} {Physical Review B}\ }\textbf
		{\bibinfo {volume} {87}},\ \bibinfo {pages} {054421} (\bibinfo {year}
		{2013})}\BibitemShut {NoStop}%
	\bibitem [{\citenamefont {Guo}\ \emph {et~al.}(2016)\citenamefont {Guo},
		\citenamefont {Herklotz}, \citenamefont {Kehlberger}, \citenamefont {Cramer},
		\citenamefont {Jakob},\ and\ \citenamefont {Kl{\"a}ui}}]{guo_thermal_2016}%
	\BibitemOpen
	\bibfield  {author} {\bibinfo {author} {\bibfnamefont {E.-J.}\ \bibnamefont
			{Guo}}, \bibinfo {author} {\bibfnamefont {A.}~\bibnamefont {Herklotz}},
		\bibinfo {author} {\bibfnamefont {A.}~\bibnamefont {Kehlberger}}, \bibinfo
		{author} {\bibfnamefont {J.}~\bibnamefont {Cramer}}, \bibinfo {author}
		{\bibfnamefont {G.}~\bibnamefont {Jakob}}, \ and\ \bibinfo {author}
		{\bibfnamefont {M.}~\bibnamefont {Kl{\"a}ui}},\ }\href {\doibase
		10.1063/1.4939625} {\bibfield  {journal} {\bibinfo  {journal} {Applied
				Physics Letters}\ }\textbf {\bibinfo {volume} {108}},\ \bibinfo {pages}
		{022403} (\bibinfo {year} {2016})}\BibitemShut {NoStop}%
	\bibitem [{\citenamefont {Niizeki}\ \emph {et~al.}(2015)\citenamefont
		{Niizeki}, \citenamefont {Kikkawa}, \citenamefont {Uchida}, \citenamefont
		{Oka}, \citenamefont {Suzuki}, \citenamefont {Yanagihara}, \citenamefont
		{Kita},\ and\ \citenamefont {Saitoh}}]{niizeki_observation_2015}%
	\BibitemOpen
	\bibfield  {author} {\bibinfo {author} {\bibfnamefont {T.}~\bibnamefont
			{Niizeki}}, \bibinfo {author} {\bibfnamefont {T.}~\bibnamefont {Kikkawa}},
		\bibinfo {author} {\bibfnamefont {K.-i.}\ \bibnamefont {Uchida}}, \bibinfo
		{author} {\bibfnamefont {M.}~\bibnamefont {Oka}}, \bibinfo {author}
		{\bibfnamefont {K.~Z.}\ \bibnamefont {Suzuki}}, \bibinfo {author}
		{\bibfnamefont {H.}~\bibnamefont {Yanagihara}}, \bibinfo {author}
		{\bibfnamefont {E.}~\bibnamefont {Kita}}, \ and\ \bibinfo {author}
		{\bibfnamefont {E.}~\bibnamefont {Saitoh}},\ }\href {\doibase
		10.1063/1.4916978} {\bibfield  {journal} {\bibinfo  {journal} {AIP Advances}\
		}\textbf {\bibinfo {volume} {5}},\ \bibinfo {pages} {053603} (\bibinfo {year}
		{2015})}\BibitemShut {NoStop}%
	\bibitem [{\citenamefont {Ramos}\ \emph {et~al.}(2013)\citenamefont {Ramos},
		\citenamefont {Kikkawa}, \citenamefont {Uchida}, \citenamefont {Adachi},
		\citenamefont {Lucas}, \citenamefont {Aguirre}, \citenamefont {Algarabel},
		\citenamefont {Morell{\'o}n}, \citenamefont {Maekawa}, \citenamefont
		{Saitoh},\ and\ \citenamefont {Ibarra}}]{ramos_observation_2013}%
	\BibitemOpen
	\bibfield  {author} {\bibinfo {author} {\bibfnamefont {R.}~\bibnamefont
			{Ramos}}, \bibinfo {author} {\bibfnamefont {T.}~\bibnamefont {Kikkawa}},
		\bibinfo {author} {\bibfnamefont {K.}~\bibnamefont {Uchida}}, \bibinfo
		{author} {\bibfnamefont {H.}~\bibnamefont {Adachi}}, \bibinfo {author}
		{\bibfnamefont {I.}~\bibnamefont {Lucas}}, \bibinfo {author} {\bibfnamefont
			{M.~H.}\ \bibnamefont {Aguirre}}, \bibinfo {author} {\bibfnamefont
			{P.}~\bibnamefont {Algarabel}}, \bibinfo {author} {\bibfnamefont
			{L.}~\bibnamefont {Morell{\'o}n}}, \bibinfo {author} {\bibfnamefont
			{S.}~\bibnamefont {Maekawa}}, \bibinfo {author} {\bibfnamefont
			{E.}~\bibnamefont {Saitoh}}, \ and\ \bibinfo {author} {\bibfnamefont {M.~R.}\
			\bibnamefont {Ibarra}},\ }\href {\doibase 10.1063/1.4793486} {\bibfield
		{journal} {\bibinfo  {journal} {Applied Physics Letters}\ }\textbf {\bibinfo
			{volume} {102}},\ \bibinfo {pages} {072413} (\bibinfo {year}
		{2013})}\BibitemShut {NoStop}%
	\bibitem [{\citenamefont {Uchida}\ \emph
		{et~al.}(2010{\natexlab{b}})\citenamefont {Uchida}, \citenamefont {Nonaka},
		\citenamefont {Ota},\ and\ \citenamefont
		{Saitoh}}]{uchida_longitudinal_2010}%
	\BibitemOpen
	\bibfield  {author} {\bibinfo {author} {\bibfnamefont {K.-i.}\ \bibnamefont
			{Uchida}}, \bibinfo {author} {\bibfnamefont {T.}~\bibnamefont {Nonaka}},
		\bibinfo {author} {\bibfnamefont {T.}~\bibnamefont {Ota}}, \ and\ \bibinfo
		{author} {\bibfnamefont {E.}~\bibnamefont {Saitoh}},\ }\href {\doibase
		10.1063/1.3533397} {\bibfield  {journal} {\bibinfo  {journal} {Applied
				Physics Letters}\ }\textbf {\bibinfo {volume} {97}},\ \bibinfo {pages}
		{262504} (\bibinfo {year} {2010}{\natexlab{b}})}\BibitemShut {NoStop}%
	\bibitem [{\citenamefont {Aqeel}\ \emph {et~al.}(2015)\citenamefont {Aqeel},
		\citenamefont {Vlietstra}, \citenamefont {Heuver}, \citenamefont {Bauer},
		\citenamefont {Noheda}, \citenamefont {van Wees},\ and\ \citenamefont
		{Palstra}}]{aqeel_spin-hall_2015}%
	\BibitemOpen
	\bibfield  {author} {\bibinfo {author} {\bibfnamefont {A.}~\bibnamefont
			{Aqeel}}, \bibinfo {author} {\bibfnamefont {N.}~\bibnamefont {Vlietstra}},
		\bibinfo {author} {\bibfnamefont {J.~A.}\ \bibnamefont {Heuver}}, \bibinfo
		{author} {\bibfnamefont {G.~E.~W.}\ \bibnamefont {Bauer}}, \bibinfo {author}
		{\bibfnamefont {B.}~\bibnamefont {Noheda}}, \bibinfo {author} {\bibfnamefont
			{B.~J.}\ \bibnamefont {van Wees}}, \ and\ \bibinfo {author} {\bibfnamefont
			{T.~T.~M.}\ \bibnamefont {Palstra}},\ }\href {\doibase
		10.1103/PhysRevB.92.224410} {\bibfield  {journal} {\bibinfo  {journal}
			{Physical Review B}\ }\textbf {\bibinfo {volume} {92}},\ \bibinfo {pages}
		{224410} (\bibinfo {year} {2015})}\BibitemShut {NoStop}%
	\bibitem [{\citenamefont {Bougiatioti}\ \emph {et~al.}(2017)\citenamefont
		{Bougiatioti}, \citenamefont {Klewe}, \citenamefont {Meier}, \citenamefont
		{Manos}, \citenamefont {Kuschel}, \citenamefont {Wollschl{\"a}ger},
		\citenamefont {Bouchenoire}, \citenamefont {Brown}, \citenamefont
		{Schmalhorst}, \citenamefont {Reiss},\ and\ \citenamefont
		{Kuschel}}]{bougiatioti_quantitative_2017-1}%
	\BibitemOpen
	\bibfield  {author} {\bibinfo {author} {\bibfnamefont {P.}~\bibnamefont
			{Bougiatioti}}, \bibinfo {author} {\bibfnamefont {C.}~\bibnamefont {Klewe}},
		\bibinfo {author} {\bibfnamefont {D.}~\bibnamefont {Meier}}, \bibinfo
		{author} {\bibfnamefont {O.}~\bibnamefont {Manos}}, \bibinfo {author}
		{\bibfnamefont {O.}~\bibnamefont {Kuschel}}, \bibinfo {author} {\bibfnamefont
			{J.}~\bibnamefont {Wollschl{\"a}ger}}, \bibinfo {author} {\bibfnamefont
			{L.}~\bibnamefont {Bouchenoire}}, \bibinfo {author} {\bibfnamefont {S.~D.}\
			\bibnamefont {Brown}}, \bibinfo {author} {\bibfnamefont {J.-M.}\ \bibnamefont
			{Schmalhorst}}, \bibinfo {author} {\bibfnamefont {G.}~\bibnamefont {Reiss}},
		\ and\ \bibinfo {author} {\bibfnamefont {T.}~\bibnamefont {Kuschel}},\ }\href
	{\doibase 10.1103/PhysRevLett.119.227205} {\bibfield  {journal} {\bibinfo
			{journal} {Physical Review Letters}\ }\textbf {\bibinfo {volume} {119}},\
		\bibinfo {pages} {227205} (\bibinfo {year} {2017})}\BibitemShut {NoStop}%
	\bibitem [{\citenamefont {Kuschel}\ \emph {et~al.}(2016)\citenamefont
		{Kuschel}, \citenamefont {Klewe}, \citenamefont {Bougiatioti}, \citenamefont
		{Kuschel}, \citenamefont {Wollschl{\"a}ger}, \citenamefont {Bouchenoire},
		\citenamefont {Brown}, \citenamefont {Schmalhorst}, \citenamefont {Meier},\
		and\ \citenamefont {Reiss}}]{kuschel_static_2016}%
	\BibitemOpen
	\bibfield  {author} {\bibinfo {author} {\bibfnamefont {T.}~\bibnamefont
			{Kuschel}}, \bibinfo {author} {\bibfnamefont {C.}~\bibnamefont {Klewe}},
		\bibinfo {author} {\bibfnamefont {P.}~\bibnamefont {Bougiatioti}}, \bibinfo
		{author} {\bibfnamefont {O.}~\bibnamefont {Kuschel}}, \bibinfo {author}
		{\bibfnamefont {J.}~\bibnamefont {Wollschl{\"a}ger}}, \bibinfo {author}
		{\bibfnamefont {L.}~\bibnamefont {Bouchenoire}}, \bibinfo {author}
		{\bibfnamefont {S.~D.}\ \bibnamefont {Brown}}, \bibinfo {author}
		{\bibfnamefont {J.~M.}\ \bibnamefont {Schmalhorst}}, \bibinfo {author}
		{\bibfnamefont {D.}~\bibnamefont {Meier}}, \ and\ \bibinfo {author}
		{\bibfnamefont {G.}~\bibnamefont {Reiss}},\ }\href {\doibase
		10.1109/TMAG.2015.2512040} {\bibfield  {journal} {\bibinfo  {journal} {IEEE
				Transactions on Magnetics}\ }\textbf {\bibinfo {volume} {52}},\ \bibinfo
		{pages} {1} (\bibinfo {year} {2016})}\BibitemShut {NoStop}%
	\bibitem [{\citenamefont {Margulies}\ \emph {et~al.}(1997)\citenamefont
		{Margulies}, \citenamefont {Parker}, \citenamefont {Rudee}, \citenamefont
		{Spada}, \citenamefont {Chapman}, \citenamefont {Aitchison},\ and\
		\citenamefont {Berkowitz}}]{margulies_origin_1997}%
	\BibitemOpen
	\bibfield  {author} {\bibinfo {author} {\bibfnamefont {D.~T.}\ \bibnamefont
			{Margulies}}, \bibinfo {author} {\bibfnamefont {F.~T.}\ \bibnamefont
			{Parker}}, \bibinfo {author} {\bibfnamefont {M.~L.}\ \bibnamefont {Rudee}},
		\bibinfo {author} {\bibfnamefont {F.~E.}\ \bibnamefont {Spada}}, \bibinfo
		{author} {\bibfnamefont {J.~N.}\ \bibnamefont {Chapman}}, \bibinfo {author}
		{\bibfnamefont {P.~R.}\ \bibnamefont {Aitchison}}, \ and\ \bibinfo {author}
		{\bibfnamefont {A.~E.}\ \bibnamefont {Berkowitz}},\ }\href {\doibase
		10.1103/PhysRevLett.79.5162} {\bibfield  {journal} {\bibinfo  {journal}
			{Physical Review Letters}\ }\textbf {\bibinfo {volume} {79}},\ \bibinfo
		{pages} {5162} (\bibinfo {year} {1997})}\BibitemShut {NoStop}%
	\bibitem [{\citenamefont {Singh}\ \emph {et~al.}(2017)\citenamefont {Singh},
		\citenamefont {Khodadadi}, \citenamefont {Mohammadi}, \citenamefont
		{Keshavarz}, \citenamefont {Mewes}, \citenamefont {Negi}, \citenamefont
		{Datta}, \citenamefont {Galazka}, \citenamefont {Uecker},\ and\ \citenamefont
		{Gupta}}]{singh_bulk_2017}%
	\BibitemOpen
	\bibfield  {author} {\bibinfo {author} {\bibfnamefont {A.~V.}\ \bibnamefont
			{Singh}}, \bibinfo {author} {\bibfnamefont {B.}~\bibnamefont {Khodadadi}},
		\bibinfo {author} {\bibfnamefont {J.~B.}\ \bibnamefont {Mohammadi}}, \bibinfo
		{author} {\bibfnamefont {S.}~\bibnamefont {Keshavarz}}, \bibinfo {author}
		{\bibfnamefont {T.}~\bibnamefont {Mewes}}, \bibinfo {author} {\bibfnamefont
			{D.~S.}\ \bibnamefont {Negi}}, \bibinfo {author} {\bibfnamefont
			{R.}~\bibnamefont {Datta}}, \bibinfo {author} {\bibfnamefont
			{Z.}~\bibnamefont {Galazka}}, \bibinfo {author} {\bibfnamefont
			{R.}~\bibnamefont {Uecker}}, \ and\ \bibinfo {author} {\bibfnamefont
			{A.}~\bibnamefont {Gupta}},\ }\href {\doibase 10.1002/adma.201701222}
	{\bibfield  {journal} {\bibinfo  {journal} {Advanced Materials}\ }\textbf
		{\bibinfo {volume} {29}},\ \bibinfo {pages} {1701222} (\bibinfo {year}
		{2017})}\BibitemShut {NoStop}%
		\bibitem [{\citenamefont {Rastogi}\ \emph {et~al.}()\citenamefont {Rastogi},
		\citenamefont {Singh}, \citenamefont {Li}, \citenamefont {Peters}, \citenamefont {P. Bougiatioti}, \citenamefont {Meier},
		\citenamefont {Mohammadi}, \citenamefont {Khodadadi}, \citenamefont {Mewes},
		\citenamefont {Mishra}, \citenamefont {Gazquez}, \citenamefont {Borisevich},
		\citenamefont {Galazka}, \citenamefont {Uecker}, \citenamefont {Reiss},
		\citenamefont {Kuschel},\ and\ \citenamefont
		{Gupta}}]{rastogi_enhancement_nodate}%
	\BibitemOpen
	\bibfield  {author} {\bibinfo {author} {\bibfnamefont {A.}~\bibnamefont
			{Rastogi}}, \bibinfo {author} {\bibfnamefont {A.~V.}\ \bibnamefont {Singh}},
		\bibinfo {author} {\bibfnamefont {Z.}~\bibnamefont {Li}}, 	\bibinfo {author} {\bibfnamefont {T.}~\bibnamefont {Peters}},	\bibinfo {author} {\bibfnamefont {P.}~\bibnamefont {Bougiatioti}}, \bibinfo {author}
		{\bibfnamefont {D.}~\bibnamefont {Meier}}, \bibinfo {author} {\bibfnamefont
			{J.~B.}\ \bibnamefont {Mohammadi}}, \bibinfo {author} {\bibfnamefont
			{B.}~\bibnamefont {Khodadadi}}, \bibinfo {author} {\bibfnamefont
			{T.}~\bibnamefont {Mewes}}, \bibinfo {author} {\bibfnamefont
			{R.}~\bibnamefont {Mishra}}, \bibinfo {author} {\bibfnamefont
			{J.}~\bibnamefont {Gazquez}}, \bibinfo {author} {\bibfnamefont {A.~Y.}\
			\bibnamefont {Borisevich}}, \bibinfo {author} {\bibfnamefont
			{Z.}~\bibnamefont {Galazka}}, \bibinfo {author} {\bibfnamefont
			{R.}~\bibnamefont {Uecker}}, \bibinfo {author} {\bibfnamefont
			{G.}~\bibnamefont {Reiss}}, \bibinfo {author} {\bibfnamefont
			{T.}~\bibnamefont {Kuschel}}, \ and\ \bibinfo {author} {\bibfnamefont
			{A.}~\bibnamefont {Gupta}},\ }\href@noop {} {\bibinfo  {journal} {submitted}\
	}\BibitemShut {NoStop}%
	\bibitem [{\citenamefont {Schreier}\ \emph {et~al.}(2013)\citenamefont
		{Schreier}, \citenamefont {Roschewsky}, \citenamefont {Dobler}, \citenamefont
		{Meyer}, \citenamefont {Huebl}, \citenamefont {Gross},\ and\ \citenamefont
		{Goennenwein}}]{schreier_current_2013}%
	\BibitemOpen
	\bibfield  {journal} {  }\bibfield  {author} {\bibinfo {author} {\bibfnamefont
			{M.}~\bibnamefont {Schreier}}, \bibinfo {author} {\bibfnamefont
			{N.}~\bibnamefont {Roschewsky}}, \bibinfo {author} {\bibfnamefont
			{E.}~\bibnamefont {Dobler}}, \bibinfo {author} {\bibfnamefont
			{S.}~\bibnamefont {Meyer}}, \bibinfo {author} {\bibfnamefont
			{H.}~\bibnamefont {Huebl}}, \bibinfo {author} {\bibfnamefont
			{R.}~\bibnamefont {Gross}}, \ and\ \bibinfo {author} {\bibfnamefont
			{S.~T.~B.}\ \bibnamefont {Goennenwein}},\ }\href {\doibase 10.1063/1.4839395}
	{\bibfield  {journal} {\bibinfo  {journal} {Applied Physics Letters}\
		}\textbf {\bibinfo {volume} {103}},\ \bibinfo {pages} {242404} (\bibinfo
		{year} {2013})}\BibitemShut {NoStop}%
	\bibitem [{\citenamefont {Vlietstra}\ \emph {et~al.}(2014)\citenamefont
		{Vlietstra}, \citenamefont {Shan}, \citenamefont {van Wees}, \citenamefont
		{Isasa}, \citenamefont {Casanova},\ and\ \citenamefont
		{Ben~Youssef}}]{vlietstra_simultaneous_2014}%
	\BibitemOpen
	\bibfield  {author} {\bibinfo {author} {\bibfnamefont {N.}~\bibnamefont
			{Vlietstra}}, \bibinfo {author} {\bibfnamefont {J.}~\bibnamefont {Shan}},
		\bibinfo {author} {\bibfnamefont {B.~J.}\ \bibnamefont {van Wees}}, \bibinfo
		{author} {\bibfnamefont {M.}~\bibnamefont {Isasa}}, \bibinfo {author}
		{\bibfnamefont {F.}~\bibnamefont {Casanova}}, \ and\ \bibinfo {author}
		{\bibfnamefont {J.}~\bibnamefont {Ben~Youssef}},\ }\href {\doibase
		10.1103/PhysRevB.90.174436} {\bibfield  {journal} {\bibinfo  {journal}
			{Physical Review B}\ }\textbf {\bibinfo {volume} {90}},\ \bibinfo {pages}
		{174436} (\bibinfo {year} {2014})}\BibitemShut {NoStop}%
	\bibitem [{\citenamefont {Shan}\ \emph {et~al.}(2016)\citenamefont {Shan},
		\citenamefont {Cornelissen}, \citenamefont {Vlietstra}, \citenamefont
		{Ben~Youssef}, \citenamefont {Kuschel}, \citenamefont {Duine},\ and\
		\citenamefont {van Wees}}]{shan_influence_2016}%
	\BibitemOpen
	\bibfield  {author} {\bibinfo {author} {\bibfnamefont {J.}~\bibnamefont
			{Shan}}, \bibinfo {author} {\bibfnamefont {L.~J.}\ \bibnamefont
			{Cornelissen}}, \bibinfo {author} {\bibfnamefont {N.}~\bibnamefont
			{Vlietstra}}, \bibinfo {author} {\bibfnamefont {J.}~\bibnamefont
			{Ben~Youssef}}, \bibinfo {author} {\bibfnamefont {T.}~\bibnamefont
			{Kuschel}}, \bibinfo {author} {\bibfnamefont {R.~A.}\ \bibnamefont {Duine}},
		\ and\ \bibinfo {author} {\bibfnamefont {B.~J.}\ \bibnamefont {van Wees}},\
	}\href {\doibase 10.1103/PhysRevB.94.174437} {\bibfield  {journal} {\bibinfo
			{journal} {Physical Review B}\ }\textbf {\bibinfo {volume} {94}},\ \bibinfo
		{pages} {174437} (\bibinfo {year} {2016})}\BibitemShut {NoStop}%
	\bibitem [{\citenamefont {Giles}\ \emph {et~al.}(2015)\citenamefont {Giles},
		\citenamefont {Yang}, \citenamefont {Jamison},\ and\ \citenamefont
		{Myers}}]{giles_long-range_2015}%
	\BibitemOpen
	\bibfield  {author} {\bibinfo {author} {\bibfnamefont {B.~L.}\ \bibnamefont
			{Giles}}, \bibinfo {author} {\bibfnamefont {Z.}~\bibnamefont {Yang}},
		\bibinfo {author} {\bibfnamefont {J.~S.}\ \bibnamefont {Jamison}}, \ and\
		\bibinfo {author} {\bibfnamefont {R.~C.}\ \bibnamefont {Myers}},\ }\href
	{\doibase 10.1103/PhysRevB.92.224415} {\bibfield  {journal} {\bibinfo
			{journal} {Physical Review B}\ }\textbf {\bibinfo {volume} {92}},\ \bibinfo
		{pages} {224415} (\bibinfo {year} {2015})}\BibitemShut {NoStop}%
	\bibitem [{\citenamefont {Shan}\ \emph
		{et~al.}(2017{\natexlab{b}})\citenamefont {Shan}, \citenamefont
		{Cornelissen}, \citenamefont {Liu}, \citenamefont {Youssef}, \citenamefont
		{Liang},\ and\ \citenamefont {van Wees}}]{shan_criteria_2017}%
	\BibitemOpen
	\bibfield  {author} {\bibinfo {author} {\bibfnamefont {J.}~\bibnamefont
			{Shan}}, \bibinfo {author} {\bibfnamefont {L.~J.}\ \bibnamefont
			{Cornelissen}}, \bibinfo {author} {\bibfnamefont {J.}~\bibnamefont {Liu}},
		\bibinfo {author} {\bibfnamefont {J.~B.}\ \bibnamefont {Youssef}}, \bibinfo
		{author} {\bibfnamefont {L.}~\bibnamefont {Liang}}, \ and\ \bibinfo {author}
		{\bibfnamefont {B.~J.}\ \bibnamefont {van Wees}},\ }\href {\doibase
		10.1103/PhysRevB.96.184427} {\bibfield  {journal} {\bibinfo  {journal}
			{Physical Review B}\ }\textbf {\bibinfo {volume} {96}},\ \bibinfo {pages}
		{184427} (\bibinfo {year} {2017}{\natexlab{b}})}\BibitemShut {NoStop}%
	\bibitem [{\citenamefont {Giles}\ \emph {et~al.}(2017)\citenamefont {Giles},
		\citenamefont {Yang}, \citenamefont {Jamison}, \citenamefont {Gomez-Perez},
		\citenamefont {V{\'e}lez}, \citenamefont {Hueso}, \citenamefont {Casanova},\
		and\ \citenamefont {Myers}}]{giles_thermally_2017}%
	\BibitemOpen
	\bibfield  {author} {\bibinfo {author} {\bibfnamefont {B.~L.}\ \bibnamefont
			{Giles}}, \bibinfo {author} {\bibfnamefont {Z.}~\bibnamefont {Yang}},
		\bibinfo {author} {\bibfnamefont {J.~S.}\ \bibnamefont {Jamison}}, \bibinfo
		{author} {\bibfnamefont {J.~M.}\ \bibnamefont {Gomez-Perez}}, \bibinfo
		{author} {\bibfnamefont {S.}~\bibnamefont {V{\'e}lez}}, \bibinfo {author}
		{\bibfnamefont {L.~E.}\ \bibnamefont {Hueso}}, \bibinfo {author}
		{\bibfnamefont {F.}~\bibnamefont {Casanova}}, \ and\ \bibinfo {author}
		{\bibfnamefont {R.~C.}\ \bibnamefont {Myers}},\ }\href {\doibase
		10.1103/PhysRevB.96.180412} {\bibfield  {journal} {\bibinfo  {journal}
			{Physical Review B}\ }\textbf {\bibinfo {volume} {96}},\ \bibinfo {pages}
		{180412} (\bibinfo {year} {2017})}\BibitemShut {NoStop}%
	\bibitem [{\citenamefont {Cornelissen}\ \emph {et~al.}(2017)\citenamefont
		{Cornelissen}, \citenamefont {Oyanagi}, \citenamefont {Kikkawa},
		\citenamefont {Qiu}, \citenamefont {Kuschel}, \citenamefont {Bauer},
		\citenamefont {van Wees},\ and\ \citenamefont
		{Saitoh}}]{cornelissen_nonlocal_2017}%
	\BibitemOpen
	\bibfield  {author} {\bibinfo {author} {\bibfnamefont {L.~J.}\ \bibnamefont
			{Cornelissen}}, \bibinfo {author} {\bibfnamefont {K.}~\bibnamefont
			{Oyanagi}}, \bibinfo {author} {\bibfnamefont {T.}~\bibnamefont {Kikkawa}},
		\bibinfo {author} {\bibfnamefont {Z.}~\bibnamefont {Qiu}}, \bibinfo {author}
		{\bibfnamefont {T.}~\bibnamefont {Kuschel}}, \bibinfo {author} {\bibfnamefont
			{G.~E.~W.}\ \bibnamefont {Bauer}}, \bibinfo {author} {\bibfnamefont {B.~J.}\
			\bibnamefont {van Wees}}, \ and\ \bibinfo {author} {\bibfnamefont
			{E.}~\bibnamefont {Saitoh}},\ }\href {\doibase 10.1103/PhysRevB.96.104441}
	{\bibfield  {journal} {\bibinfo  {journal} {Physical Review B}\ }\textbf
		{\bibinfo {volume} {96}},\ \bibinfo {pages} {104441} (\bibinfo {year}
		{2017})}\BibitemShut {NoStop}%
	\bibitem [{\citenamefont {V{\'e}lez}\ \emph {et~al.}(2016)\citenamefont
		{V{\'e}lez}, \citenamefont {Golovach}, \citenamefont {Bedoya-Pinto},
		\citenamefont {Isasa}, \citenamefont {Sagasta}, \citenamefont {Abadia},
		\citenamefont {Rogero}, \citenamefont {Hueso}, \citenamefont {Bergeret},\
		and\ \citenamefont {Casanova}}]{velez_hanle_2016}%
	\BibitemOpen
	\bibfield  {author} {\bibinfo {author} {\bibfnamefont {S.}~\bibnamefont
			{V{\'e}lez}}, \bibinfo {author} {\bibfnamefont {V.~N.}\ \bibnamefont
			{Golovach}}, \bibinfo {author} {\bibfnamefont {A.}~\bibnamefont
			{Bedoya-Pinto}}, \bibinfo {author} {\bibfnamefont {M.}~\bibnamefont {Isasa}},
		\bibinfo {author} {\bibfnamefont {E.}~\bibnamefont {Sagasta}}, \bibinfo
		{author} {\bibfnamefont {M.}~\bibnamefont {Abadia}}, \bibinfo {author}
		{\bibfnamefont {C.}~\bibnamefont {Rogero}}, \bibinfo {author} {\bibfnamefont
			{L.~E.}\ \bibnamefont {Hueso}}, \bibinfo {author} {\bibfnamefont {F.~S.}\
			\bibnamefont {Bergeret}}, \ and\ \bibinfo {author} {\bibfnamefont
			{F.}~\bibnamefont {Casanova}},\ }\href {\doibase
		10.1103/PhysRevLett.116.016603} {\bibfield  {journal} {\bibinfo  {journal}
			{Physical Review Letters}\ }\textbf {\bibinfo {volume} {116}},\ \bibinfo
		{pages} {016603} (\bibinfo {year} {2016})}\BibitemShut {NoStop}%
	\bibitem [{\citenamefont {Vlietstra}\ \emph {et~al.}(2013)\citenamefont
		{Vlietstra}, \citenamefont {Shan}, \citenamefont {Castel}, \citenamefont {van
			Wees},\ and\ \citenamefont {Ben~Youssef}}]{vlietstra_spin-hall_2013}%
	\BibitemOpen
	\bibfield  {author} {\bibinfo {author} {\bibfnamefont {N.}~\bibnamefont
			{Vlietstra}}, \bibinfo {author} {\bibfnamefont {J.}~\bibnamefont {Shan}},
		\bibinfo {author} {\bibfnamefont {V.}~\bibnamefont {Castel}}, \bibinfo
		{author} {\bibfnamefont {B.~J.}\ \bibnamefont {van Wees}}, \ and\ \bibinfo
		{author} {\bibfnamefont {J.}~\bibnamefont {Ben~Youssef}},\ }\href {\doibase
		10.1103/PhysRevB.87.184421} {\bibfield  {journal} {\bibinfo  {journal}
			{Physical Review B}\ }\textbf {\bibinfo {volume} {87}},\ \bibinfo {pages}
		{184421} (\bibinfo {year} {2013})}\BibitemShut {NoStop}%
	\bibitem [{\citenamefont {Chen}\ \emph {et~al.}(2013)\citenamefont {Chen},
		\citenamefont {Takahashi}, \citenamefont {Nakayama}, \citenamefont
		{Althammer}, \citenamefont {Goennenwein}, \citenamefont {Saitoh},\ and\
		\citenamefont {Bauer}}]{chen_theory_2013}%
	\BibitemOpen
	\bibfield  {author} {\bibinfo {author} {\bibfnamefont {Y.-T.}\ \bibnamefont
			{Chen}}, \bibinfo {author} {\bibfnamefont {S.}~\bibnamefont {Takahashi}},
		\bibinfo {author} {\bibfnamefont {H.}~\bibnamefont {Nakayama}}, \bibinfo
		{author} {\bibfnamefont {M.}~\bibnamefont {Althammer}}, \bibinfo {author}
		{\bibfnamefont {S.~T.~B.}\ \bibnamefont {Goennenwein}}, \bibinfo {author}
		{\bibfnamefont {E.}~\bibnamefont {Saitoh}}, \ and\ \bibinfo {author}
		{\bibfnamefont {G.~E.~W.}\ \bibnamefont {Bauer}},\ }\href {\doibase
		10.1103/PhysRevB.87.144411} {\bibfield  {journal} {\bibinfo  {journal}
			{Physical Review B}\ }\textbf {\bibinfo {volume} {87}},\ \bibinfo {pages}
		{144411} (\bibinfo {year} {2013})}\BibitemShut {NoStop}%
	\bibitem [{\citenamefont {Liu}\ \emph {et~al.}(2017)\citenamefont {Liu},
		\citenamefont {Cornelissen}, \citenamefont {Shan}, \citenamefont {Kuschel},\
		and\ \citenamefont {van Wees}}]{liu_magnon_2017}%
	\BibitemOpen
	\bibfield  {author} {\bibinfo {author} {\bibfnamefont {J.}~\bibnamefont
			{Liu}}, \bibinfo {author} {\bibfnamefont {L.~J.}\ \bibnamefont
			{Cornelissen}}, \bibinfo {author} {\bibfnamefont {J.}~\bibnamefont {Shan}},
		\bibinfo {author} {\bibfnamefont {T.}~\bibnamefont {Kuschel}}, \ and\
		\bibinfo {author} {\bibfnamefont {B.~J.}\ \bibnamefont {van Wees}},\ }\href
	{\doibase 10.1103/PhysRevB.95.140402} {\bibfield  {journal} {\bibinfo
			{journal} {Physical Review B}\ }\textbf {\bibinfo {volume} {95}},\ \bibinfo
		{pages} {140402} (\bibinfo {year} {2017})}\BibitemShut {NoStop}%
	\bibitem [{\citenamefont {Liu}\ \emph {et~al.}(2018)\citenamefont {Liu},
		\citenamefont {Cornelissen}, \citenamefont {Shan}, \citenamefont {Wees},\
		and\ \citenamefont {Kuschel}}]{liu_nonlocal_2018}%
	\BibitemOpen
	\bibfield  {author} {\bibinfo {author} {\bibfnamefont {J.}~\bibnamefont
			{Liu}}, \bibinfo {author} {\bibfnamefont {L.~J.}\ \bibnamefont
			{Cornelissen}}, \bibinfo {author} {\bibfnamefont {J.}~\bibnamefont {Shan}},
		\bibinfo {author} {\bibfnamefont {B.~J.~v.}\ \bibnamefont {Wees}}, \ and\
		\bibinfo {author} {\bibfnamefont {T.}~\bibnamefont {Kuschel}},\ }\href
	{\doibase 10.1088/1361-6463/aabf80} {\bibfield  {journal} {\bibinfo
			{journal} {Journal of Physics D: Applied Physics}\ }\textbf {\bibinfo
			{volume} {51}},\ \bibinfo {pages} {224005} (\bibinfo {year}
		{2018})}\BibitemShut {NoStop}%
	\bibitem [{\citenamefont {Duine}\ \emph {et~al.}(2017)\citenamefont {Duine},
		\citenamefont {Brataas}, \citenamefont {Bender},\ and\ \citenamefont
		{Tserkovnyak}}]{duine_universal_2017}%
	\BibitemOpen
	\bibfield  {author} {\bibinfo {author} {\bibfnamefont {R.~A.}\ \bibnamefont
			{Duine}}, \bibinfo {author} {\bibfnamefont {A.}~\bibnamefont {Brataas}},
		\bibinfo {author} {\bibfnamefont {S.~A.}\ \bibnamefont {Bender}}, \ and\
		\bibinfo {author} {\bibfnamefont {Y.}~\bibnamefont {Tserkovnyak}},\ }\href
	{http://www.cambridge.org/nl/academic/subjects/physics/condensed-matter-physics-nanoscience-and-mesoscopic-physics/universal-themes-bose-einstein-condensation}
	{\emph {\bibinfo {title} {Universal themes of {Bose}-{Einstein} condensation,
				chapter 26}}}\ (\bibinfo  {publisher} {Cambridge University Press},\ \bibinfo
	{address} {Cambridge, United Kingdom},\ \bibinfo {year} {2017})\ \bibinfo
	{note} {edited by David Snoke, Nikolaos Proukakis and Peter
		Littlewood}\BibitemShut {NoStop}%
	\bibitem [{\citenamefont {Kikkawa}\ \emph {et~al.}(2016)\citenamefont
		{Kikkawa}, \citenamefont {Shen}, \citenamefont {Flebus}, \citenamefont
		{Duine}, \citenamefont {Uchida}, \citenamefont {Qiu}, \citenamefont {Bauer},\
		and\ \citenamefont {Saitoh}}]{kikkawa_magnon_2016}%
	\BibitemOpen
	\bibfield  {author} {\bibinfo {author} {\bibfnamefont {T.}~\bibnamefont
			{Kikkawa}}, \bibinfo {author} {\bibfnamefont {K.}~\bibnamefont {Shen}},
		\bibinfo {author} {\bibfnamefont {B.}~\bibnamefont {Flebus}}, \bibinfo
		{author} {\bibfnamefont {R.~A.}\ \bibnamefont {Duine}}, \bibinfo {author}
		{\bibfnamefont {K.-i.}\ \bibnamefont {Uchida}}, \bibinfo {author}
		{\bibfnamefont {Z.}~\bibnamefont {Qiu}}, \bibinfo {author} {\bibfnamefont
			{G.~E.~W.}\ \bibnamefont {Bauer}}, \ and\ \bibinfo {author} {\bibfnamefont
			{E.}~\bibnamefont {Saitoh}},\ }\href {\doibase
		10.1103/PhysRevLett.117.207203} {\bibfield  {journal} {\bibinfo  {journal}
			{Physical Review Letters}\ }\textbf {\bibinfo {volume} {117}},\ \bibinfo
		{pages} {207203} (\bibinfo {year} {2016})}\BibitemShut {NoStop}%
	\bibitem [{\citenamefont {Flebus}\ \emph {et~al.}(2017)\citenamefont {Flebus},
		\citenamefont {Shen}, \citenamefont {Kikkawa}, \citenamefont {Uchida},
		\citenamefont {Qiu}, \citenamefont {Saitoh}, \citenamefont {Duine},\ and\
		\citenamefont {Bauer}}]{flebus_magnon-polaron_2017}%
	\BibitemOpen
	\bibfield  {author} {\bibinfo {author} {\bibfnamefont {B.}~\bibnamefont
			{Flebus}}, \bibinfo {author} {\bibfnamefont {K.}~\bibnamefont {Shen}},
		\bibinfo {author} {\bibfnamefont {T.}~\bibnamefont {Kikkawa}}, \bibinfo
		{author} {\bibfnamefont {K.-i.}\ \bibnamefont {Uchida}}, \bibinfo {author}
		{\bibfnamefont {Z.}~\bibnamefont {Qiu}}, \bibinfo {author} {\bibfnamefont
			{E.}~\bibnamefont {Saitoh}}, \bibinfo {author} {\bibfnamefont {R.~A.}\
			\bibnamefont {Duine}}, \ and\ \bibinfo {author} {\bibfnamefont {G.~E.~W.}\
			\bibnamefont {Bauer}},\ }\href {\doibase 10.1103/PhysRevB.95.144420}
	{\bibfield  {journal} {\bibinfo  {journal} {Physical Review B}\ }\textbf
		{\bibinfo {volume} {95}},\ \bibinfo {pages} {144420} (\bibinfo {year}
		{2017})}\BibitemShut {NoStop}%
	\bibitem [{\citenamefont {Man}\ \emph {et~al.}(2017)\citenamefont {Man},
		\citenamefont {Shi}, \citenamefont {Xu}, \citenamefont {Xu}, \citenamefont
		{Chen}, \citenamefont {Sullivan}, \citenamefont {Zhou}, \citenamefont {Xia},
		\citenamefont {Shi},\ and\ \citenamefont {Dai}}]{man_direct_2017}%
	\BibitemOpen
	\bibfield  {author} {\bibinfo {author} {\bibfnamefont {H.}~\bibnamefont
			{Man}}, \bibinfo {author} {\bibfnamefont {Z.}~\bibnamefont {Shi}}, \bibinfo
		{author} {\bibfnamefont {G.}~\bibnamefont {Xu}}, \bibinfo {author}
		{\bibfnamefont {Y.}~\bibnamefont {Xu}}, \bibinfo {author} {\bibfnamefont
			{X.}~\bibnamefont {Chen}}, \bibinfo {author} {\bibfnamefont {S.}~\bibnamefont
			{Sullivan}}, \bibinfo {author} {\bibfnamefont {J.}~\bibnamefont {Zhou}},
		\bibinfo {author} {\bibfnamefont {K.}~\bibnamefont {Xia}}, \bibinfo {author}
		{\bibfnamefont {J.}~\bibnamefont {Shi}}, \ and\ \bibinfo {author}
		{\bibfnamefont {P.}~\bibnamefont {Dai}},\ }\href {\doibase
		10.1103/PhysRevB.96.100406} {\bibfield  {journal} {\bibinfo  {journal}
			{Physical Review B}\ }\textbf {\bibinfo {volume} {96}},\ \bibinfo {pages}
		{100406} (\bibinfo {year} {2017})}\BibitemShut {NoStop}%
	\bibitem [{\citenamefont {Wang}\ \emph {et~al.}(2018)\citenamefont {Wang},
		\citenamefont {Hou}, \citenamefont {Kikkawa}, \citenamefont {Ramos},
		\citenamefont {Shen}, \citenamefont {Qiu}, \citenamefont {Chen},
		\citenamefont {Umeda}, \citenamefont {Shiomi}, \citenamefont {Jin},\ and\
		\citenamefont {Saitoh}}]{wang_bimodal_2018}%
	\BibitemOpen
	\bibfield  {author} {\bibinfo {author} {\bibfnamefont {H.}~\bibnamefont
			{Wang}}, \bibinfo {author} {\bibfnamefont {D.}~\bibnamefont {Hou}}, \bibinfo
		{author} {\bibfnamefont {T.}~\bibnamefont {Kikkawa}}, \bibinfo {author}
		{\bibfnamefont {R.}~\bibnamefont {Ramos}}, \bibinfo {author} {\bibfnamefont
			{K.}~\bibnamefont {Shen}}, \bibinfo {author} {\bibfnamefont {Z.}~\bibnamefont
			{Qiu}}, \bibinfo {author} {\bibfnamefont {Y.}~\bibnamefont {Chen}}, \bibinfo
		{author} {\bibfnamefont {M.}~\bibnamefont {Umeda}}, \bibinfo {author}
		{\bibfnamefont {Y.}~\bibnamefont {Shiomi}}, \bibinfo {author} {\bibfnamefont
			{X.}~\bibnamefont {Jin}}, \ and\ \bibinfo {author} {\bibfnamefont
			{E.}~\bibnamefont {Saitoh}},\ }\href {\doibase 10.1063/1.5022195} {\bibfield
		{journal} {\bibinfo  {journal} {Applied Physics Letters}\ }\textbf {\bibinfo
			{volume} {112}},\ \bibinfo {pages} {142406} (\bibinfo {year}
		{2018})}\BibitemShut {NoStop}%
	\bibitem [{\citenamefont {G{\"u}lseren}\ and\ \citenamefont
		{Cohen}(2002)}]{gulseren_high-pressure_2002}%
	\BibitemOpen
	\bibfield  {author} {\bibinfo {author} {\bibfnamefont {O.}~\bibnamefont
			{G{\"u}lseren}}\ and\ \bibinfo {author} {\bibfnamefont {R.~E.}\ \bibnamefont
			{Cohen}},\ }\href {\doibase 10.1103/PhysRevB.65.064103} {\bibfield  {journal}
		{\bibinfo  {journal} {Physical Review B}\ }\textbf {\bibinfo {volume} {65}},\
		\bibinfo {pages} {064103} (\bibinfo {year} {2002})}\BibitemShut {NoStop}%
	\bibitem [{\citenamefont {Li}\ and\ \citenamefont
		{Fisher}(1990)}]{li_single_1990}%
	\BibitemOpen
	\bibfield  {author} {\bibinfo {author} {\bibfnamefont {Z.}~\bibnamefont
			{Li}}\ and\ \bibinfo {author} {\bibfnamefont {E.~S.}\ \bibnamefont
			{Fisher}},\ }\href {\doibase 10.1007/BF00720147} {\bibfield  {journal}
		{\bibinfo  {journal} {Journal of Materials Science Letters}\ }\textbf
		{\bibinfo {volume} {9}},\ \bibinfo {pages} {759} (\bibinfo {year}
		{1990})}\BibitemShut {NoStop}%
	\bibitem [{\citenamefont {Srivastava}\ and\ \citenamefont
		{Aiyar}(1987)}]{srivastava_spin_1987}%
	\BibitemOpen
	\bibfield  {author} {\bibinfo {author} {\bibfnamefont {C.~M.}\ \bibnamefont
			{Srivastava}}\ and\ \bibinfo {author} {\bibfnamefont {R.}~\bibnamefont
			{Aiyar}},\ }\href {\doibase 10.1088/0022-3719/20/8/013} {\bibfield  {journal}
		{\bibinfo  {journal} {Journal of Physics C: Solid State Physics}\ }\textbf
		{\bibinfo {volume} {20}},\ \bibinfo {pages} {1119} (\bibinfo {year}
		{1987})}\BibitemShut {NoStop}%
	\bibitem [{\citenamefont {Franco}, \citenamefont {Pessoni},\ and\ \citenamefont
		{Machado}(2015)}]{franco_spin-wave_2015}%
	\BibitemOpen
	\bibfield  {author} {\bibinfo {author} {\bibfnamefont {A.}~\bibnamefont
			{Franco}}, \bibinfo {author} {\bibfnamefont {H.~V.~S.}\ \bibnamefont
			{Pessoni}}, \ and\ \bibinfo {author} {\bibfnamefont {F.~L.~A.}\ \bibnamefont
			{Machado}},\ }\href {\doibase 10.1063/1.4934749} {\bibfield  {journal}
		{\bibinfo  {journal} {Journal of Applied Physics}\ }\textbf {\bibinfo
			{volume} {118}},\ \bibinfo {pages} {173904} (\bibinfo {year}
		{2015})}\BibitemShut {NoStop}%
	\bibitem [{\citenamefont {Kodama}\ \emph {et~al.}(1996)\citenamefont {Kodama},
		\citenamefont {Berkowitz}, \citenamefont {McNiff},\ and\ \citenamefont
		{Foner}}]{kodama_surface_1996}%
	\BibitemOpen
	\bibfield  {author} {\bibinfo {author} {\bibfnamefont {R.~H.}\ \bibnamefont
			{Kodama}}, \bibinfo {author} {\bibfnamefont {A.~E.}\ \bibnamefont
			{Berkowitz}}, \bibinfo {author} {\bibfnamefont {J.}~\bibnamefont {McNiff},
			\bibfnamefont {E.~J.}}, \ and\ \bibinfo {author} {\bibfnamefont
			{S.}~\bibnamefont {Foner}},\ }\href {\doibase 10.1103/PhysRevLett.77.394}
	{\bibfield  {journal} {\bibinfo  {journal} {Physical Review Letters}\
		}\textbf {\bibinfo {volume} {77}},\ \bibinfo {pages} {394} (\bibinfo {year}
		{1996})}\BibitemShut {NoStop}%
\end{thebibliography}
\end{document}